\documentclass[12pt, draftcls, onecolumn]{IEEEtran}
\usepackage{setspace}
\usepackage{amsmath,amsfonts}
\usepackage{algorithmic}
\usepackage{algorithm}
\usepackage{array}
\usepackage[caption=false,font=normalsize,labelfont=scriptsize,textfont=scriptsize]{subfig}
\usepackage{textcomp}
\usepackage{stfloats}
\usepackage{url}
\usepackage{verbatim}
\usepackage{graphicx}
\usepackage{subfig}
\usepackage{cite}
\usepackage{amssymb}
\usepackage{siunitx}
\usepackage{threeparttable}
\usepackage{bm}
\begin{document}

\title{Environment-Aware Codebook Design for RIS-Assisted MU-MISO Communications: \\Implementation and Performance Analysis}

\author{Zhiheng Yu, Jiancheng An,~\IEEEmembership{Member,~IEEE,} \\Ertugrul Basar,~\IEEEmembership{Fellow,~IEEE}, Lu Gan, and Chau Yuen,~\IEEEmembership{Fellow,~IEEE}
\thanks{This article was presented in part at the IEEE VTC2024-Spring \cite{arXiv_Yu_2024_Environment}. This work was partially supported by Sichuan Science and Technology Program under Grant 2023YFSY0008 and 2023YFG0291. The work of Ertugrul Basar was supported by TUBITAK under Grant 120E401. The work of Chau Yuen was supported in part by the Ministry of Education, Singapore, under its Ministry of Education (MOE) Tier 2 under Award MOE-T2EP50220-0019; and in part by the Science and Engineering Research Council of Agency for Science, Technology and Research (A*STAR) Singapore, under Grant M22L1b0110.

Z. Yu and L. Gan are with the national key laboratory on blind signal processing. School of Information and Communication Engineering, University of Electronic Science and Technology of China (UESTC), Chengdu 611731, China, and also with the Yibin Institute of UESTC, Yibin 644000, China (e-mail: zhihengyu2000@163.com; ganlu@uestc.edu.cn).

J. An and C. Yuen are with the School of Electrical and Electronics Engineering, Nanyang Technological University, Singapore 639798 (e-mail: jiancheng\underline{~}an@163.com; chau.yuen@ntu.edu.sg). 

E. Basar is with the Communications Research and Innovation Laboratory (CoreLab), Department of Electrical and Electronics Engineering, Koç University, Sariyer, Istanbul 34450, Turkey. (e-mail: ebasar@ku.edu.tr). }}

\markboth{DRAFT}%
{Shell \MakeLowercase{\textit{et al.}}: A Sample Article Using IEEEtran.cls for IEEE Journals}

\maketitle

\begin{abstract}
Reconfigurable intelligent surface (RIS) provides a new electromagnetic response control solution, which can proactively reshape the characteristics of wireless channel environments. In RIS-assisted communication systems, the acquisition of channel state information (CSI) and the optimization of reflecting coefficients constitute major design challenges. To address these issues, codebook-based solutions have been developed recently, which, however, are mostly environment-agnostic. In this paper, a novel environment-aware codebook protocol is proposed, which can significantly reduce both pilot overhead and computational complexity, while maintaining expected communication performance. Specifically, first of all, a channel training framework is introduced to divide the training phase into several blocks. In each block, we directly estimate the composite end-to-end channel and focus only on the transmit beamforming. Second, we propose an environment-aware codebook generation scheme, which first generates a group of channels based on statistical CSI, and then obtains their corresponding RIS configuration by utilizing the alternating optimization (AO) method offline. In each online training block, the RIS is configured based on the corresponding codeword in the environment-aware codebook, and the optimal codeword resulting in the highest sum rate is adopted for assisting in the downlink data transmission. Third, we analyze the theoretical performance of the environment-aware codebook-based protocol taking into account the channel estimation errors. Finally, numerical simulations are provided to verify our theoretical analysis and the performance of the proposed scheme. In particular, the simulation results demonstrate that our protocol is more competitive than conventional environment-agnostic codebooks.
\end{abstract}

\begin{IEEEkeywords}
Reconfigurable intelligent surface (RIS), channel training, environment-aware codebook, multiuser downlink beamforming.
\end{IEEEkeywords}

\section{Introduction}
\IEEEPARstart{R}{econfigurable} intelligent surface (RIS) is an artificial electromagnetic (EM) material with programmable EM properties \cite{r1}. Essentially, an RIS is composed of numerous sophisticatedly designed EM meta-atoms. By dynamically adjusting the control signals for each RIS element, it becomes possible to intelligently manipulate the EM characteristics, including the amplitude, phase, polarization, and frequency of the incident signal in a programming manner \cite{r2},\cite{more1}. In contrast to conventional wireless technologies such as massive multiple-input multiple-output (MIMO) and millimeter wave (mmWave) communications \cite{add2} that have to grapple with issues related to network energy consumption and hardware costs, the RIS elements are passive and simply reflect incident signals, mitigating the power-thirsty radio frequency (RF) chains and forward delay in traditional relays \cite{r3,r4,shi2023ris}. Remarkably, an RIS can operate in full-duplex mode without encountering self-interference problems \cite{r5}. By adjusting the phase shifts at the RIS, various functionalities such as secure communications, focused beam emission, and enhanced indoor coverage can be achieved \cite{add5,add4,more3}. In addition to the single-layer RIS, the application of stacked intelligent metasurfaces (SIM) in different scenarios has also begun to emerge for performing signal processing tasks in the wave domain \cite{WCM_2024_An_Stacked, SIM1,SIM2}. Besides, integrating RIS with transceivers also attracts increased attention\cite{HMIMO}. Given these advantages, RIS-empowered communication is considered to be a key technology enabling future wireless networks \cite{r1,add1,r6,shi2022spatially,shi2023uplink}.

Nevertheless, RIS-assisted wireless communication systems also present some challenges. On the one hand, obtaining accurate channel state information (CSI) is much more difficult due to the fact that most RISs are composed of passive elements or only equipped with a small number of onboard signal processing units \cite{r7}, \cite{r8}. More specifically, it becomes impractical to separately estimate base station (BS)-RIS and RIS-user channels using traditional pilot-based methods. Therefore, several advanced channel estimation techniques have been proposed to estimate cascaded or composite end-to-end channels \cite{r9,r10,r16,on/off_mmse,add7,add8,more4,DL_CE}. For instance, \emph{Mishra et al.} introduced a simple ON/OFF method capable of estimating the cascaded BS-RIS-user channels one by one \cite{r16}. Subsequently, \emph{Xu et al.} proposed a novel ON/OFF based minimum mean square error (MMSE) channel estimation method for RIS-assisted high mobility systems, which mitigates the Doppler-induced error floor\cite{on/off_mmse}. In order to reduce the pilot overhead required for estimating the cascaded BS-RIS-user channels, \emph{Wang et al.} leveraged the similarity of the BS-RIS link between multiple users \cite{r10}. Furthermore, in \cite{add7}, \emph{Wei et al.} considered the double-structured sparsity of the cascaded channels to estimate the angle parameters, while \cite{r9} proposed a channel training-based scheme to directly estimate the composite channels. More recently, \emph{Guo et al.} developed a new channel estimator leveraging the unitary approximate message passing (UAMP) technique \cite{add8}, and \emph{Chen et al.} proposed a new compressed sensing-based channel estimation protocol to estimate the cascaded channels\cite{more4, TWC_2023_An_Fundamental}. Combined with deep learning, \emph{Xu et al.} proposed a three-stage joint channel decomposition and prediction framework that exploits the quasi-static property of BS-RIS channel and the fast time-varying property of RIS-user channels\cite{DL_CE}.

On the other hand, the optimization of reflecting coefficients (RCs) at the RIS constitutes another challenge \cite{r13,r15,r19,add9,add10,add11,add12,DRL_MIMO,more6}. More specifically, in \cite{r19}, \emph{Wu and Zhang} concentrated on minimizing the total transmit power under specific signal-to-interference-plus-noise ratio (SINR) constraints, they adopted alternating optimization (AO) method based on semidefinite relaxation (SDR) to obtain approximate solutions of transmit precoding matrix and RIS RCs. In \cite{r13}, \emph{Yan et al.} devised RCs based on the statistical CSI to maximize spectral efficiency. In \cite{add11}, \emph{Di et al.} proposed a hybrid beamforming scheme to maximize the sum rate of RIS-aided multi-user multiple-input single-output (MU-MISO) systems. For beyond-diagonal (BD) RIS-aided multi-user systems, \emph{Fang and Mao} proposed a closed-form solution for passive beamforming to maximize the sum of the effective channel gains \cite{more6}. Besides, \cite{add10} proposed a heuristic algorithm to optimize continuous phase shifts of RIS-aided massive MIMO systems. In an RIS-aided mmWave MIMO system, \emph{Xu et al.} developed a novel deep reinforcement learning (DRL) algorithm to design joint beamforming according to a location-aware imitation environment\cite{DRL_MIMO}. However, it is important to note that the schemes mentioned above assume the perfect CSI, which is generally unavailable in practical scenarios. Considering the imperfect CSI, in \cite{r15}, \emph{An et al.} conceived a low-complexity framework for maximizing the achievable rate of RIS-aided MIMO system with discrete phase shifts. The RCs are pre-designed and the effective composite channel is estimated. Moreover, \cite{add9} considered an imperfect CSI scenario and solved the max-min SINR problem considering a continuous phase shift. In \cite{add12}, \emph{Nikolaos et al.} proposed two phase shift design schemes whose complexity is linear to the number of RIS elements and receive antennas.

\begin{table*}[t]
\scriptsize
\renewcommand\arraystretch{1.25}
 \begin{center}
 \caption{Comparison of Proposed Scheme with Recent RIS Solutions}
 \label{tab:table1}
 \begin{threeparttable}
 \begin{tabular}{c||c|c|c|c|c|c|c}
 \hline
 \textbf{Reference} & \textbf{MIMO setup} & \textbf{Phase shift model} & \textbf{Optimization objective} & \textbf{Channel} & \textbf{Complexity} & \textbf{Imperfect CSI} & \textbf{PBF design}\\
 \hline
 Proposed & MU-MISO & Discrete & Sum rate & Composite & Low & \checkmark & \checkmark \\
 \hline
 \cite{r9} & MU-MIMO & Discrete & Transmit power & Composite & Low & \checkmark & \checkmark \\
 \hline
 \cite{r15} & MIMO & Discrete & Throughput & Composite & Low & \checkmark & \checkmark \\ 
 \hline
 \cite{r19} & MU-MISO & Continuous & Transmit power & Separate & High & $\times$ & $\times$\\ 
 \hline
 \cite{add1} & MIMO & Continuous & Throughput & Separate & High & $\times$ & \checkmark \\ 
 \hline
 \cite{add24} & MIMO & Discrete & Received signal power & Separate & Moderate & $\times$ & \checkmark\\ 
 \hline
 \cite{add20} & MISO & Discrete & Achievable rate & Composite & Low & \checkmark & \checkmark\\ 
 \hline
 \cite{add7} & MU-MISO & Continuous & NMSE & Separate & Moderate & $\times$ & $\times$ \\ 
 \hline
 \cite{add12} & MIMO & Discrete & ASE & Separate & Moderate & \checkmark & \checkmark \\ 
 \hline
 \cite{add8} & MIMO & Continuous & NMSE & Separate & Low & $\times$ & $\times$ \\ 
 \hline
 \cite{r10} & MU-MISO & Continuous & NMSE & Separate & Moderate & \checkmark & $\times$ \\ 
 \hline
 \end{tabular}
 \begin{tablenotes}
 \footnotesize
 \item ASE: Average spectral efficiency \quad NMSE: Normalized mean square error \quad PBF: Passive beamforming
 \end{tablenotes}
 \end{threeparttable}
 \end{center}
\end{table*}

\begin{figure}[t]
\centering
\includegraphics[width=10cm]{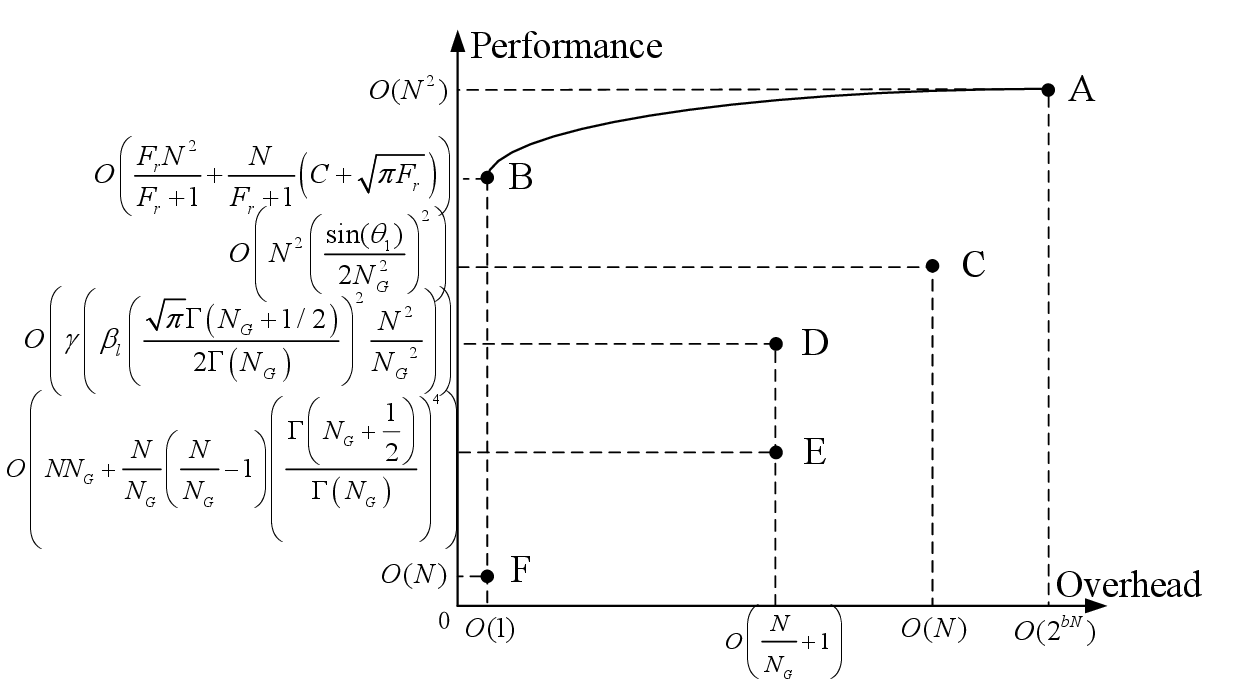}
\begin{tablenotes}
 \footnotesize
 \item A\cite{r19}: Optimal configuration \hspace{0.25cm} B\cite{add20}: Statistic CSI-based scheme
 \item C\cite{add24}: Scalar/Vector RIS design D\cite{add25}: Optimal RIS grouping 
 \item E\cite{add26}: Group connected RIS \hspace{0.32cm} F\cite{r19}: Random configuration
 \item B-A curve: Proposed environment-aware codebook scheme
\end{tablenotes}
\begin{tablenotes}
\footnotesize
\item $F_r$\hspace{0.12cm}: Rician factor of the RIS-user channel 
\item $C$\hspace{0.22cm}: Euler-Mascheroni constant
\item $\theta_1$\hspace{0.16cm}: RC angle of the reactance matrix's non-zero eigenvalues
\item $\gamma$\hspace{0.27cm}: Transmit SNR
\item $\Gamma$\hspace{0.26cm}: Gamma function
\item $N_G$: Number of RIS elements in each group
\item $\beta_l$\hspace{0.18cm}: Large-scale path loss coefficient of the cascaded link
\end{tablenotes}
\caption{ The performance versus overhead diagram, where the proposed scheme strikes a flexible tradeoff between the statistical CSI-based scheme and the optimal RIS configuration.}
\label{fig_0}
\end{figure} 

In addition, considering the complexity and overhead, \cite{ATselect} proposed a low-complexity antenna selection algorithm to avoid the joint optimization problem with the transceiver, which is capable of supporting arbitrary MIMO configuration. Moreover, several group-/fully-connected schemes are introduced to reduce the pilot overhead. More specifically, in \cite{add25}, \emph{Shen et al.} proposed a novel RIS architecture based on group- and fully-connected reconfigurable impedance networks and derived the scaling law for received signals in RIS-aided MIMO systems. Adopting a similar RIS structure, \cite{add24} analyzed the scaling law for received signal power using BD RISs in case of discrete phase shifts. Additionally, in \cite{add26}, \emph{Kundu et al.} adopted a grouping strategy to optimize the RIS and derived a tight closed-form upper bound for the achievable rate.

Recently, various codebook-based schemes have emerged to enhance RIS-based systems' performance with moderate overhead. For instance, \cite{add21} proposed a codebook-based framework and introduced three codebook generation schemes. Further in \cite{r8} and \cite{r9}, \emph{An et al.} employed a low complexity framework to maximize the achievable rate. They analyzed the theoretical performance of the random and uniform codebooks. Beyond generating codebooks offline, in \cite{add20}, \emph{Jia et al.} proposed an environment-aware codebook design to flexibly adapt to different channel conditions. Moreover, \cite{add22} proposed a discrete Fourier transform (DFT) codebook, which, however, experiences significant SINR loss in the near-field region of RIS \cite{WCM_2024_An_Near}. To solve this problem, a ring-type codebook based on the Fresnel principle was further proposed to determine the phase distribution by employing the location information of the user and an RIS. Besides, in \cite{add23}, \emph{Abdallah et al.} developed a multi-agent DRL-based beamforming codebook design that relies only on receiving power measurements.

Nevertheless, the above codebook-based schemes usually entail substantial training overhead and lack adaptability to channel environment variations. Against this background, in this paper, we introduce a novel codebook-based protocol for RIS-assisted MU-MISO systems and propose an environment-aware scheme to obtain a discrete RIS RC configuration codebook. Specifically, we first generate virtual channels in each training block according to the statistical CSI, and then perform AO of RC configuration and transmit power allocation to obtain the corresponding RIS RC codeword. In the on-line stage, we configure the RIS according to the predesigned codebook and estimate the composite end-to-end channel for each codeword, which reduces the performance loss caused by imperfect CSI. Then, we theoretically analyze the received power scaling law in a single-user scenario considering both the perfect and imperfect CSI to gain valuable insights into the proposed environment-aware codebook scheme. Comparing with the existing work \cite{add20}, we derive a tighter upper bound of the achievable rate in perfect CSI scenario. Moreover, the influence of channel estimation errors in the environment-aware codebook-based schemes has been theoretically analyzed in \emph{Proposition 2}. For illustration, we contrast the proposed scheme to its existing channel estimation and passive beamforming counterparts in Table \ref{tab:table1}. In addition, as shown in Fig. \ref{fig_0}, we show the performance vs. overhead tradeoff of the proposed scheme and contrast it to existing schemes. To summarize, our contributions are outlined as follows.

\begin{itemize}
\item[1)] We propose a channel training-based protocol for RIS-assisted MU-MISO systems. In contrast to conventional schemes, we develop a novel environment-aware codebook design scheme, where we generate a set of virtual channels with the same distribution as practical channels. For each virtual channel, we employ the AO method to solve the sum rate maximization problem. Specifically, we adopt a successive refinement method to optimize the RC at the RIS and a classic water-filling power allocation algorithm to allocate transmit power at the BS when one of them is fixed. By doing so for $Q$ virtual channels, we generate an environment-aware RC codebook.
\item[2)] Then, in the on-line stage, we configure the RIS using each codeword in the environment-aware codebook and thus obtain $Q$ candidate composite end-to-end channels. The optimal index that maximizes the sum rate is determined to obtain the corresponding discrete RIS configuration and transmit power allocation.
\item[3)] Next, the theoretical achievable rate adopting the proposed environment-aware channel training-based protocol is analyzed under both the perfect and imperfect CSI scenarios. To the best of our knowledge, this is the first work to analyze the influence of channel estimation errors on codeword index selection. 
\item[4)] Finally, numerical simulation results are provided to validate our performance analysis. Both the theoretical analysis and simulation results demonstrate that the proposed environment-aware codebook scheme strikes a flexible trade-off between the achievable performance and pilot overhead.
\end{itemize}

The rest of this paper is organized as follows. Section II introduces the RIS-assisted multi-user system model. In Section III, we introduce the proposed codebook-based protocol and environment-aware codebook generation scheme. Then, the theoretical performance of the proposed scheme is analyzed in Section IV, where the effects of channel estimation errors are also considered. Section V provides numerical simulations to evaluate the performance of the proposed scheme. Finally, Section VI concludes the paper.

$\emph{Notations:}$ Scalars are denoted by italic letters, vectors and matrices are denoted by lower and upper bold-face letters, respectively; $\mathbb{C}^{x \times y}$ denotes the space of $x \times y$ complex-valued matrices; For any general matrix $\mathbf{A}$, $\mathbf{A}^T, \mathbf{A}^H, \mathbf{A}^*, \text{rank}(\mathbf{A})$ represent its transpose, Hermitian transpose, conjugate and rank, respectively; For a complex-valued vector $\mathbf x$, $\lVert \mathbf x \rVert $ denotes its Euclidean norm, $|\mathbf x|$ denotes its magnitude, and $\text{diag}(\mathbf x)$ denotes a diagonal matrix with each diagonal element being the corresponding entry in $\mathbf x$; We denote the $N \times N$ identity matrix as $\mathbf{I}_N$; $\text{log}(\cdot)$ and $\lfloor {\cdot} \rfloor$ represent the logarithmic function and floor operator, respectively; The distribution of a circularly symmetric complex Gaussian (CSCG) random vector with mean vector $\mathbf x$ and covariance matrix $\mathbf{\Sigma}$ is denoted by $ \sim \mathcal{CN} (\mathbf x, \mathbf{\Sigma})$, where $\sim$ stands for “distributed as”; For a square matrix $\mathbf S$, $\text{tr}(\mathbf S)$ and $\mathbf {S}^{-1}$ denote its trace and inverse, respectively; $\mathbb{E}\left\{{\cdot}\right\}$ denotes the statistical expectation. $\Re \left\{{\cdot}\right\}$ and $\Im\left\{{\cdot}\right\}$ denotes the real part and imagine part of a complex number, respectively. For clarity, we summarize the main symbols used in this paper and their meanings in Table \ref{tab:table3}.

\begin{table}[!t]
\renewcommand\arraystretch{1.25}
 \begin{center} 
 \caption{Major Symbols and Their Meanings}
 \label{tab:table3}
 \begin{tabular}{c||l} 
 \hline
 Symbol & Meaning \\
 \hline
 $M$ & The number of BS antennas \\
 \hline
 $N$ & The number of RIS elements \\
 \hline
 $K$ & The number of users \\
 \hline
 $Q$ & Training overhead \\
 \hline
 $\mathbf{G}$ & The BS-RIS channel \\
 \hline
 $\mathbf{h}_{d,k}^H$ & The BS-user $k$ channel \\
 \hline
 $\mathbf{h}_{r,k}^H$ & The RIS-user $k$ channel \\
 \hline
 $\mathbf{h}_k^H$ & The composite channel of user $k$ \\
 \hline
 $\mathbf{\Phi}$ & RC configuration of the RIS \\
 \hline
 $\mathbf{w}$ & Transmit precoding vector \\
 \hline
 $\theta$ & Phase shift of RIS element \\
 \hline
 $R$ & Sum rate of multiple users \\
 \hline 
 $T_c$ & The channel coherence time \\
 \hline
 $\overline{\mathbf{P}}$ & Transmit power allocation matrix \\
 \hline
 $\varepsilon$ & Composite channel estimation error \\
 \hline
 $P_d$ & Transmit power at the BS \\
 \hline
 $P_{\text{ul}}$ & Average power of the pilot signals \\
 \hline
 $\sigma_z^2$ & Average noise power at the BS \\
 \hline
 $\sigma_k^2$ & Average noise power at the users \\ 
 \hline
 \end{tabular}
 \end{center}
\end{table}

\section{System Model}
\begin{figure}[t]
\centering
\includegraphics[width=10cm]{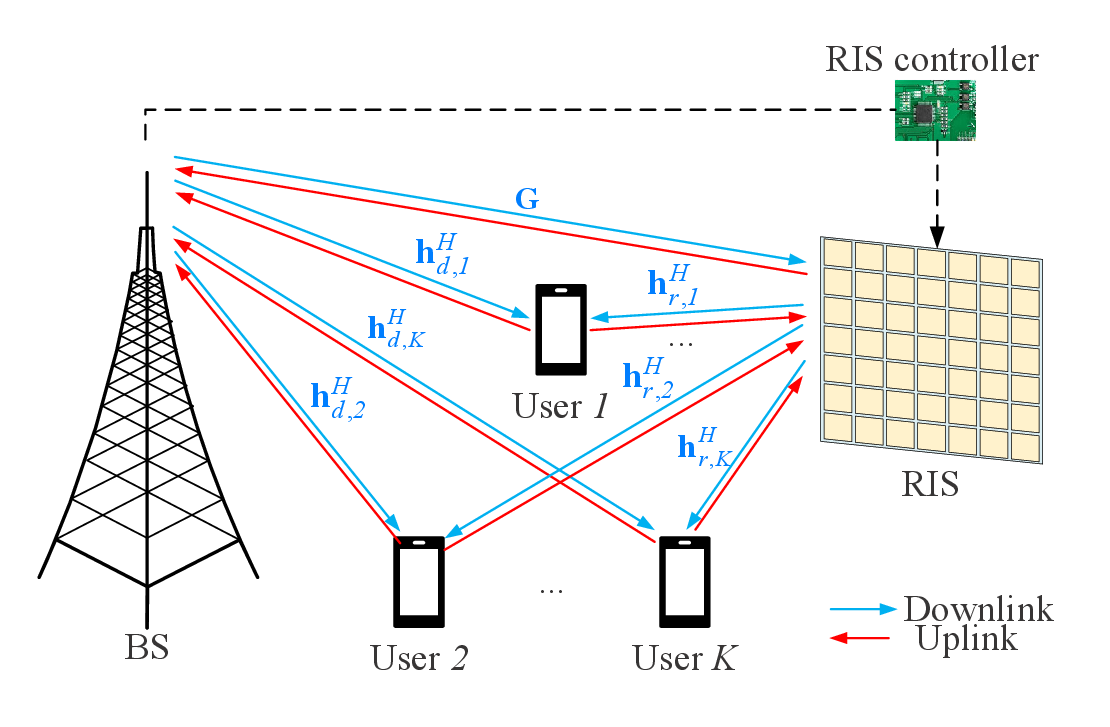}
\caption{ An RIS-assisted multiuser communication system. }
\label{fig_1}
\end{figure} 

Let us consider an RIS-assisted multi-user communication system in a single cell as shown in Fig. \ref{fig_1}, where an RIS with $N$ reflecting elements is deployed to enhance the downlink (DL) signal transmission between a multi-antenna BS and $K$ single-antenna users. The number of transmit antennas at the BS is denoted by $M$. The RIS is equipped with a smart controller capable of adjusting the RCs according to the real-time CSI \cite{r7}. In this paper, we adopt a quasi-static flat-fading channel model for all links. Besides, we consider a time-division duplexing (TDD) protocol for both uplink (UL) as well as DL transmissions and assume the channel’s reciprocity for the CSI acquisition in the DL based on the UL training. The baseband equivalent channels spanning from the BS to the RIS, from the RIS to user $k$, and from the BS to user $k$ are denoted by $\mathbf{G} \in \mathbb{C}^{N \times M}$, $\mathbf{h}_{r,k}^H \in \mathbb{C}^{1 \times N}$, $\mathbf{h}_{d,k}^H \in \mathbb{C}^{1 \times M}$, respectively, with $k = 1, 2, \cdots, K$.

Let the diagonal matrix $\mathbf{\Phi} = \text{diag} \left( \varphi_1, \varphi_2, \cdots, \varphi_N \right)$ represent the RC configuration of the RIS, where $\varphi_n$ denotes the RC of the $n$th RIS element, following $\left| \varphi_n \right| = 1$ for $n = 1,2,\cdots,N$. In this paper, we use discrete phase shifts to control signal reflection. In order to facilitate the practical implementation, we consider that the discrete phase shifts are obtained by uniformly quantizing the interval $\left[0, 2\pi \right)$ and let $b$ represent the number of bits used for quantizing the phase shift levels. Hence the phase shift of each RIS element is assumed to be one of $B =2^b$ discrete phase shift values. The phase shift set $\mathcal B$ is given by 
\begin{align} \mathcal B = \left \{0, \varDelta \theta, \cdots, (B-1) \varDelta \theta \right \}, \end{align} where $\varDelta \theta =2 \pi /B $. Thus, the RC of the $n$th element at the RIS can be expressed as $\varphi_n = e^{j \theta_n} $, $\theta_n \in \mathcal{B}$, $n = 1,2,\cdots,N$ and the composite end-to-end channel for user $k$ is given by 
\begin{align} \mathbf{h}_k^H = \mathbf{h}_{r,k}^H \mathbf{\Phi} \mathbf{G} + \mathbf{h}_{d,k}^H, \quad k = 1, 2, \cdots, K. \end{align} 
Let $\mathbf{H} = \left[ \mathbf{h}_1, \mathbf{h}_2, \cdots, \mathbf{h}_K \right] \in \mathbb{C}^{M \times K}$ represent the UL channels associated with $K$ users.

During the channel estimation process in the UL phase, the pilot signal received at the BS is given by 
\begin{align} \mathbf{y} = \sum_{k=1}^{K} \mathbf{h}_k \sqrt{P_{\text{ul}}} x_k + \mathbf{z}, \end{align} 
where $P_{\text{ul}}$ is the average power of the pilot symbols, which is assumed to be identical for all users, $x_k$ denotes the pilot symbol transmitted from user $k$ to BS with zero mean and unit variance, $\mathbf{z} \in \mathbb{C}^{M \times 1}$ denotes the additive white Gaussian noise (AWGN) at the BS obeying $\mathbf{z} \sim \mathcal{CN} \left( \mathbf{0},\sigma_z^2 \mathbf{I}_M \right)$. 

Next, we consider the DL of data transmission and use linear transmit precoding at the BS. As a result, the complex baseband signal received at user $k$ is given by 
\begin{align} r_k= \mathbf{h}_k^H \sum_{i=1}^{K} \mathbf{w}_i s_i + n_k, \end{align}
where $s_k$ denotes the transmitted information symbol of user $k$, $\mathbf{w}_k \in \mathbb{C}^{M \times 1}$ denotes the corresponding transmit beamforming vector and $n_k \sim \mathcal{CN} \left( 0,\sigma_k^2 \right), k = 1,2,\cdots,K $ denotes the AWGN at the $k$th user’s receiver.

Meanwhile, we adopt the Rician channel in this paper. Specifically, the RIS-user $k$ channel can be expressed as 
\begin{align} \label{eq:5} \mathbf{h}_{r,k} = \sqrt{\beta_r} \left( \sqrt{\frac{F_r}{F_r + 1}}\mathbf{h}_{r,k}^{\text{LoS}} + \sqrt{\frac{1}{F_r + 1}}\mathbf{h}_{r,k}^{\text{NLoS}} \right), \end{align} 
where $\beta_r$ and $F_r$ are the path loss and the Rician factor of RIS-user channel, respectively; $\mathbf{h}_{r,k}^{\text{LoS}}$ and $\mathbf{h}_{r,k}^{\text{NLoS}}$ represent the line-of-sight (LoS) and non-line-of-sight (NLoS) components of RIS-user channel, respectively. The NLoS components are i.i.d. complex Gaussian variables with zero mean and unit variance, satisfying $\mathbf{h}_{r,k}^{\text{NLoS}} \sim \mathcal{CN} \left(\mathbf{0}, \mathbf{I}_N \right)$. Similarly, the BS-user channel and BS-RIS channel are modeled by using (\ref{eq:5}).

Moreover, we consider a uniform linear array (ULA) of $M$ antennas at the BS and a uniform planar array (UPA) of $N$ antennas at the RIS. Let $\mathbf{a}_{\text{BS}}\left(\delta\right) \in \mathbb{C}^{M \times 1}$ and $\mathbf{a}_{\text{R}}\left(\zeta, \gamma\right) \in \mathbb{C}^{N \times 1}$ denote the steering vector of the BS and the RIS, respectively. Specifically, the $m$th entry of $\mathbf{a}_{\text{BS}}$ is denoted as $e^{j \frac{2 \pi}{\lambda} (m-1) d_{\text{BS}} \sin(\delta) }, m = 1, 2, \cdots, M$, where $d_{\text{BS}}$ denotes the element spacing of the BS, $\lambda$ denotes the wavelength of the signal, and $\delta \in \left[-\pi/2, \pi/2\right)$ denotes the angle of departure (AoD) or the angle of arrival (AoA). The $n$th entry of $\mathbf{a}_{\text{R}}$ is denoted as $e^{j \frac{2 \pi}{\lambda} d_{\text{R}} \sin(\gamma) \left[ \lfloor \frac{n-1}{N_x} \rfloor \sin(\zeta) + ((n-1)-\lfloor \frac{n-1}{N_x} \rfloor N_x) \cos(\zeta) \right]/\lambda }, n = 1, 2, \cdots, N$, where $d_{\text{R}}$ denotes the element spacing of the RIS. $N_x$ is the number of reflecting elements deployed at each row of the RIS. Moreover, $\zeta \in \left[0, \pi\right)$ and $\gamma \in \left[-\pi/2, \pi/2\right)$ denote the azimuth and elevation AoA/AoD, respectively. Thus, the LoS component of the $\mathbf{G}$, $\mathbf{h}_r$ and $\mathbf{h}_d$ are given by $\mathbf{a}_{\text{BS}}\left(\delta_d^{\text{AoA}}\right)$, $\mathbf{a}_{\text{R}}\left(\zeta_g^{\text{AoA}}, \gamma_g^{\text{AoA}}\right) \mathbf{a}_{\text{BS}}\left(\delta_g^{\text{AoD}}\right)^H$ and $\mathbf{a}_{\text{R}}\left(\zeta_r^{\text{AoA}}, \gamma_r^{\text{AoA}}\right)$, respectively, where $\delta_d^{\text{AoA}}$ and $\delta_g^{\text{AoD}}$ represent the AoA from the user to the BS and the AoD from the BS to the RIS, respectively; $\zeta_g^{\text{AoA}}$ and $\gamma_g^{\text{AoA}}$ represent the azimuth and elevation AoA from the BS to the RIS, respectively; $\zeta_r^{\text{AoA}}$ and $\gamma_r^{\text{AoA}}$ represent the azimuth and elevation AoA from the user to the RIS, respectively.

\section{The Proposed Channel Training Protocol}
In this section, we outline the proposed codebook-based scheme by first introducing our channel training-based protocol in Section III-A. Then, our bespoke environment-aware codebook generation scheme is proposed in Section III-B. In Section III-C, we introduce the on-line codebook configuration stage.

\subsection{Channel Training-Based Protocol}

\begin{figure}[t]
\centering
\includegraphics[width=10cm]{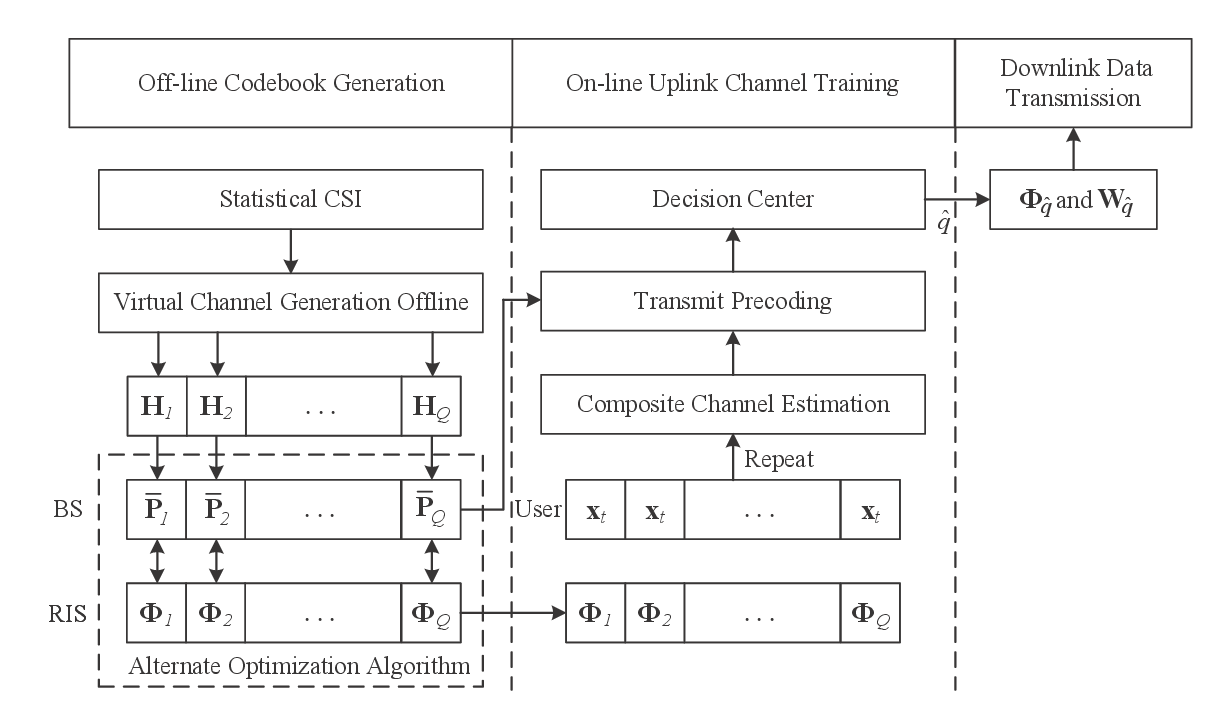}
\caption{The frame structure of the proposed protocol. }
\label{fig_framework}
\end{figure} 

As shown in Fig. \ref{fig_framework}, our channel training-based protocol consists of \emph{off-line} and \emph{on-line} stages.

In the off-line stage, we generate a group of virtual channels based on the statistical CSI. Then, we use the AO algorithm to obtain the optimal RC configuration corresponding to each virtual channel. Repeating for $Q$ virtual channels generates the RC codebook.

In the on-line stage, the transmission protocol is divided into two phases: 
\begin{itemize}
\item \emph{The UL channel training phase}: During the UL channel training phase, we adjust the RIS configuration over training blocks according to the environment-aware codebook. Then, we repeat carrying out the composite channel estimation and the transmit precoding design to obtain $Q$ candidate channels;
\item \emph{The DL data transmission phase}: During the DL signal transmission phase, we select the optimal channel maximizing the sum rate from $Q$ candidates, and get the corresponding RIS configuration and transmit power allocation to assist in the DL communications. \end{itemize}

\subsection{Off-Line Codebook Design Stage}
Next, we elaborate on the detailed steps to generate the environment-aware codebook.

\subsubsection{Virtual Channel Generation Based on Statistical CSI}
Note that the statistical CSI of all channels is easier to obtain by measurement or averaging historical observations, and it has a longer channel coherence time \cite{wang2018statistical,r13}. In this paper, we utilize the statistical CSI, which includes the path loss, the LoS component and the distribution of the NLoS component of the practical channels to design the environment-aware codebook offline. Specifically, we generate $Q$ virtual channels off-line for the RIS-user $k$ link according to (\ref{eq:5}), yielding \begin{align} \label{eq:12} \mathbf{h}_{r,k,q} = \sqrt{\beta_r} \left( \sqrt{\frac{F_r}{F_r + 1}} \mathbf{h}_{r,k}^{\text{LoS}} + \sqrt{\frac{1}{F_r + 1}} \mathbf{\tilde{h}}_{r,k,q}^{\text{NLoS}} \right), \end{align} where $\mathbf{h}_{r,k}^{\text{LoS}}$ is the LoS component of the virtual channel and it remains unchanged within each channel coherence block, $\mathbf{\tilde{h}}_{r,k,q}^{\text{NLoS}}$ represents the NLoS component of the $q$th virtual channel which is randomly generated following the same distribution with $\mathbf{h}_{r,k}^{\text{NLoS}}$ during each training block. In addition, the virtual channels for BS-RIS and BS-user $k$ links are generated in the same way.

\subsubsection{Alternating Optimization Algorithm}
For each tentative virtual channel, we endeavor to maximize the sum rate for all users by jointly optimizing the transmit precoding at the BS and the RC configuration at the RIS, subject to the total transmit power. Specifically, the optimization problem is formulated as 
\begin{align} \label{eq:9} \max_{\mathbf{W}_q, \mathbf{\Phi}_q} & \sum_{k=1}^{K} \text{log}_2 \left( 1 + \frac{\left| \mathbf{h}_{k,q}^H \mathbf{w}_{k,q} \right|^2}{\sum_{l \neq k}^K \left| \mathbf{h}_{k,q}^H \mathbf{w}_{l,q} \right|^2 + \sigma_k^2} \right) \notag\\ \mathrm{s.t.} & \mathbf{\Phi}_q = \text{diag} \left( \varphi_{q,1}, \varphi_{q,2}, \cdots, \varphi_{q,N} \right), \notag \\ & \varphi_{q,n} = e^{j \theta_{q,n}},\quad \theta_{q,n} \in \mathcal B,\quad \forall{n \in \mathcal{N}}, \notag\\& \sum_{k=1}^{K} \lVert \mathbf{w}_{k,q}\rVert^2 \leq P_d, \end{align} 
for $q = 1, 2, \cdots, Q$, where $\mathbf{h}_{k,q}^H = \mathbf{h}_{r,k}^H \mathbf{\Phi}_q \mathbf{G} + \mathbf{h}_{d,k}^H \in \mathbb{C}^{1 \times M}$ and $\mathbf{w}_{k,q}$ are the composite channel and the transmit precoding vector for user $k$ during the $q$th training block, respectively. $\mathbf{W}_q = \left[\mathbf{w}_{1,q}, \mathbf{w}_{2,q}, \cdots, \mathbf{w}_{K,q} \right] \in \mathbb{C}^{M \times K}$ denotes the transmit precoding matrix. $P_d$ is the total transmit power at the BS.

In this paper, we adopt zero-forcing (ZF) beamforming at the BS to eliminate interference between different users, yielding \begin{align} \mathbf{h}_{k,q}^H \mathbf{w}_{k,q} &= \sqrt{p_{k,q}}, k = 1, 2, \cdots, K, \notag\\ \mathbf{h}_{k,q}^H \mathbf{w}_{l,q} &= 0, \forall l \neq k, l = 1, 2, \cdots, K, \end{align} where $p_{k,q}$ denotes the received power of user $k$ in the $q$th training block.

Let $\mathbf{H}_q^H = \left[\mathbf{h}_{1,q}, \mathbf{h}_{2,q}, \cdots, \mathbf{h}_{K,q} \right]^H \in \mathbb{C}^{K \times M}, q = 1, 2, \cdots, Q$ represent the DL composite channels in the $q$th training block, the optimal transmit precoding matrix $\mathbf{W}_q$ is given by a pseudo-inverse of $\mathbf{H}_q^H$ with an appropriate power allocation among users. More specifically, the transmit precoding matrix can be expressed as 
\begin{align} \label{eq:w} \mathbf{W}_{q} = \mathbf{H}_{q} \left(\mathbf{H}_{q}^H \mathbf{H}_{q} \right)^{-1} \mathbf{P}_q^{\frac{1}{2}}, \end{align} 
where $\mathbf{P}_q = \text{diag}\left(p_{1,q},p_{2,q},\cdots ,p_{K,q} \right)$ denotes the received power matrix in the $q$th training block.

As a result, the sum rate $R_q$ can be simplified as
\begin{align} \label{eq:13} R_q = \sum_{k=1}^{K} \text{log}_2 \left(1 + \frac{p_{k,q}}{\sigma_k^2} \right). \end{align} 

Next, we utilize the AO algorithm to obtain the optimal transmit power allocation and RC solution, which consists of the following two parts:

\paragraph{Transmit Power Allocation}

Given a tentative RIS RC configuration matrix $\mathbf{\Phi}_q$, the transmit power allocated to user $k$ is $\overline{p}_{k,q} = \left \Vert \mathbf{w}_{k,q} \right \Vert^2$. 
Let $\mathbf{U}_q = \left( \mathbf{H}_q^H \mathbf{H}_q \right)^{-1} \in \mathbb{C}^{K \times K}$, $\overline{\mathbf{P}}_q = \text{diag}\left(\overline{p}_{1,q}, \overline{p}_{2,q}, \cdots, \overline{p}_{K,q} \right)$ denotes the transmit power allocation matrix. By applying the classic water-filling power allocation algorithm \cite{r15}, the transmit power allocated to the $k$th user can be expressed as \begin{align} \label{eq:pk} \overline{p}_{k,q} = \text{max} \left\{ \frac{1}{\eta_q} - \sigma_k^2 u_{k,q}, 0 \right\},\end{align} where $u_{k,q}$ is the $k$th diagonal element of $\mathbf{U}_q$ and $\eta_q$ is a threshold that satisfies $\sum_{k=1}^{K} \text{max} \left\{ \frac{1}{\eta_q} - \sigma_k^2 u_{k,q}, 0 \right\} = P_d $. Once the transmit power allocation among users is obtained, the received power of user $k$ can be expressed as $p_{k,q} = \overline{p}_{k,q}/u_{k,q}.$

\paragraph{RIS RC Optimization}

Given the transmit power allocation matrix $\overline{\mathbf{P}}_q$, problem (\ref{eq:9}) is reduced to
\begin{align} \label{eq:15} \max_{\theta_{q,n}} &\enspace R_q\left( \theta \right) = \sum_{k=1}^{K} \text{log}_2 \left( 1 + \frac{\overline{p}_{k,q}}{u_{k,q} \sigma_k^2} \right) \notag\\ \mathrm{s.t.} &\enspace \theta_{q,n} \in \mathcal{B} , \forall n \in \mathcal{N}. \end{align}
In this paper, we adopt a successive refinement method to optimize RIS RCs one by one. Specifically, when fixing other $(N-1)$ RCs, we calculate the objective function value $R_q\left( \theta \right)$ for all potential values of $\theta_{q,n}$ in $\mathcal B$ and choose the phase shift corresponding to the maximum value. After repeating this one-dimensional search process several times, we obtain the optimal RIS configuration $\mathbf{\Phi}_q$ corresponding to the transmit power allocation $\overline{\mathbf{P}}_q$.

\begin{algorithm}[!t]
\caption{Environment-Aware Codebook Design.}\label{alg:alg1}
\renewcommand{\algorithmicrequire}{\textbf{Input:}}
\renewcommand{\algorithmicensure}{\textbf{Output:}}
\begin{algorithmic}[1]
\REQUIRE Statistical CSI.
\ENSURE $\mathbf{\Phi}_q$ and $\overline{\mathbf{P}}_q,$ $q = 1, 2, \cdots, Q$.
\FOR{$q \in \left\{1, 2, \cdots, Q\right\}$}
\STATE Generate $\mathbf{h}_{r,k,q}$, $\mathbf{h}_{d,k,q}$, $k = 1, 2, \cdots, K$ and $\mathbf{G}_q$ according to (\ref{eq:12}).
\STATE Set $r = 1$.
\STATE Initialize the random RC matrix $\mathbf{\Phi}_q^r$.
\REPEAT
\STATE Optimize $\overline{\mathbf{P}}_q^r$ according to (\ref{eq:pk}).
\FOR{$n \in \left\{1, 2, \cdots, N\right\}$}
\STATE Optimize $\varphi_{q,n}^r = e^{j \theta_{q,n}^r}$ according to (\ref{eq:15}).
\ENDFOR
\STATE Update $r = r + 1$.
\UNTIL{The fractional decrease of $R_q$ is below a threshold $\epsilon > 0$.}
\STATE $\mathbf{\Phi}_q = \mathbf{\Phi}_q^r$ and $\overline{\mathbf{P}}_q = \overline{\mathbf{P}}_q^r$.
\ENDFOR
\RETURN $\mathbf{\Phi}_q$ and $\overline{\mathbf{P}}_q$.
\end{algorithmic}
\label{alg1}
\end{algorithm}

Moreover, we use random phase shifts to initialize power allocation. Then we alternately optimize the power allocation and RIS configuration until the objective function $R_q$ reaches convergence.

By carrying out the virtual channel generation in (\ref{eq:12}) and the AO method for all $Q$ training blocks, we obtain $Q$ candidate channels and their corresponding $\overline{\mathbf{P}}_q$ and $\mathbf{\Phi}_q$. Note that our channels are generated offline according to the statistical CSI, the corresponding RIS configuration and power allocation are also performed offline. 

\subsection{On-Line RIS Configuration Stage}

In the on-line stage, we complete the RIS configuration by selecting the optimal codeword in the environment-aware codebook to assist in the DL transmission. Specifically, we adjust the RIS RC in different training blocks and obtain $Q$ objective function values. The decision center will select the optimal codeword maximizing the sum rate to assist in the DL data transmission. Accordingly, the transmit precoding matrix is also determined.

\subsubsection{Composite Channel Estimation}

Specifically, given the RC configuration of the $q$th training block $\mathbf{\Phi}_q$, $q = 1,2,\cdots,Q$ the received signal at the $t$th time slot (TS) of the $q$th training block is given by 
\begin{align} \mathbf{y}_{q,t} &= \sum_{k=1}^{K} \mathbf{h}_{k,q} \sqrt{P_{\text{ul}}} x_{k,t} + \mathbf{z}_{q,t} \notag\\ &= \sqrt{P_{\text{ul}}} \mathbf{H}_q \mathbf{x}_t + \mathbf{z}_{q,t}, \end{align}
where $\mathbf{h}_{k,q}$ denotes the composite channel spanning from the $k$th user to the BS in the $q$th training block, $x_{k,t}$ is the pilot symbol transmitted from user $k$ to the BS in TS $t$, which remains identical for all $Q$ training blocks, $\mathbf{x}_t = [x_{1,t}, x_{2,t}, \cdots, x_{K,t}]^T$ is the pilot vector of the $t$th TS. 

Upon collecting the signals of $T$ TSs in the $q$th training block, the signal received at the BS in this block is given by 
\begin{align} \mathbf{Y}_q = \left[ \mathbf{y}_{q,1}, \mathbf{y}_{q,2}, \cdots, \mathbf{y}_{q,T} \right] = \sqrt{P_{\text{ul}}} \mathbf{H}_q \mathbf{X} + \mathbf{Z}_q, \end{align} 
where $\mathbf{X} = \left[ \mathbf{x}_1, \mathbf{x}_2, \cdots, \mathbf{x}_T \right] \in \mathbb{C}^{K \times T}$ is the pilot matrix used for estimating the composite channels spanning from the $K$ users to the BS. $\mathbf{Z}_q = \left[ \mathbf{z}_{q,1}, \mathbf{z}_{q,2}, \cdots, \mathbf{z}_{q,T} \right] \in \mathbb{C}^{M \times T}$ denotes the noise matrix at the BS during the $q$th training block.

The composite channel estimation during each block is equivalent to the channel estimation for conventional MU-MISO systems. We employ the mutually orthogonal pilot design to mitigate the interference between multiple users. In order to generate $K$ groups of mutually orthogonal pilot sequences for separating the pilot signals transmitted by $K$ users, the length of the pilot sequence $T$ must satisfy $T \ge K$. Furthermore, to minimize the pilot overhead, we consider $T = K$ \cite{r17}. The least squares (LS) estimate of the composite channel $\mathbf{H}_q$ of the $q$th training block can be expressed as \begin{align} \label{eq:8}\widetilde{\mathbf{H}}_{q} = \frac{1}{K \sqrt{P_{\text{ul}}}} \mathbf{Y}_q \mathbf{X}^H. \end{align}

\subsubsection{Transmit Precoding}

Based on the estimate of the composite channel in the $q$th training block, BS performs multi-user transmit precoding by utilizing the ZF precoder: \begin{align} \label{eq:22} \widetilde{\mathbf{W}}_{q} = \widetilde{\mathbf{H}}_{q} \left(\widetilde{\mathbf{H}}_{q}^H \widetilde{\mathbf{H}}_{q} \right)^{-1} \widetilde{\mathbf{P}}_q^{\frac{1}{2}}, \end{align} where $\widetilde{\mathbf{P}}_q = \text{diag}\left(\widetilde{p}_{1,q}, \widetilde{p}_{2,q}, \cdots, \widetilde{p}_{K,q}\right)$ is the received power allocation matrix and $\widetilde{p}_{k,q} = \frac{\overline{p}_{k,q}}{\widetilde{u}_{k,q}}$ is the received power of user $k$ when configuring the $q$th codeword, $\widetilde{u}_{k,q}$ is the $k$th diagonal element of $\widetilde{\mathbf{U}}_q = \left( \widetilde{\mathbf{H}}_q^H \widetilde{\mathbf{H}}_q \right)^{-1}$.

\subsubsection{Optimal Index Selection}

Then, after obtaining $Q$ candidate channels and their corresponding transmit precoding matrices, the RIS RC configuration problem can be expressed as
\begin{align} \label{eq:11} \max_q & \enspace R_q = \sum_{k=1}^{K} \text{log}_2 \left(1 + \frac{1}{\sigma_k^2} \left|\widetilde{\mathbf{h}}_{k,q}^H \widetilde{\mathbf{w}}_{k,q} \right|^2 \right) \notag\\ \mathrm{s.t.} & \enspace 1 \leq q \leq Q .\end{align} 
Once the index $\hat{q}$ of the optimal training block is obtained, we get the RIS RC configuration $\hat{\mathbf{\Phi}} = \mathbf{\Phi}_{\hat{q}}$, and the corresponding transmit precoding matrix accordingly\footnote{Note that the RIS configuration is completed by selecting the optimal codeword that maximizes the sum rate. To further boost the performance, one can design appropriate codeword weighting methods to obtain better RIS configuration, which constitutes our future work.}.

\section{Theoretical Analysis of the Proposed Scheme}
In this section, we analyze the performance of the proposed environment-aware scheme by considering a single user case. We first analyze the theoretical achievable rate considering the noiseless channels in Section IV-A. Then, in Section IV-B, we analyze the influence of channel estimation errors on the proposed channel training scheme. 

\subsection{Perfect CSI}
Firstly, we examine the scenario where there is no channel estimation error at each training block. For brevity, we assume that the direct BS-user link is blocked and the BS-RIS channel remains only the LoS component which is likely to happen in practical scenarios. Based on the above assumptions, the theoretical received power scaling law is summarized in $\emph{Proposition 1}$.

$\emph{Proposition 1}$: Assume $h_{r,n}$ following the Rician channel model with Rician factor of $F_r$ , $n = 1, 2, \cdots, N$. For $N \gg 1$, the average signal power received by the user is given by
\begin{align} \label{eq:20} P_r = & P_d \beta_r \beta_g M N \left(F_1^2 N + F_2^2 \left(\text{log}Q+C\right) + \sqrt{\pi} F_1 F_2 \right), \end{align} 
where $P_d$ is the total transmit power of BS, $Q$ is training overhead, $ F_1 = \sqrt{\frac{F_r}{F_r+1}}$, $F_2 = \sqrt{\frac{1}{F_r+1}}$ and $C\approx0.57722$ is the Euler-Mascheroni constant.

$\emph{Proof}$: Please refer to Appendix A.

\begin{figure}[!t] 
\centering 
\subfloat[\scriptsize Perfect CSI] 
{
\label{fig_used:subfig1}\includegraphics[width=0.4\textwidth]{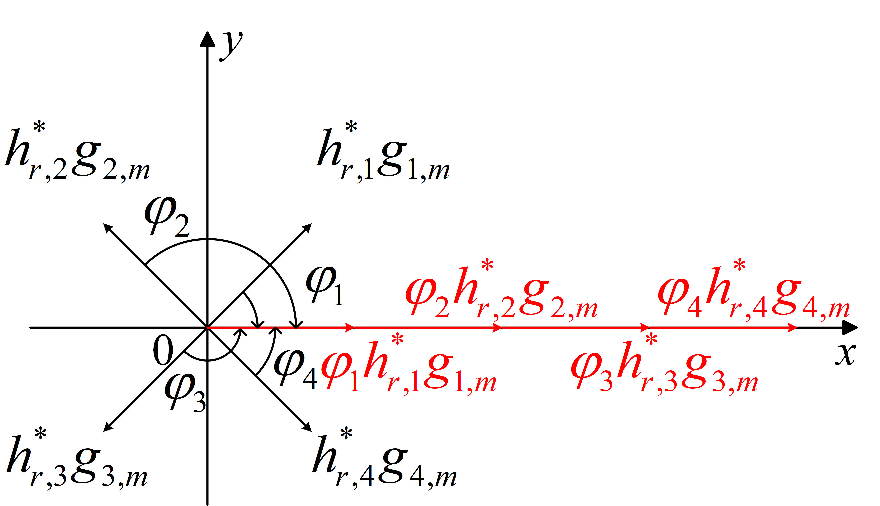}
}
\subfloat[\scriptsize Imperfect CSI]
{
\label{fig_used:subfig2}\includegraphics[width=0.4\textwidth]{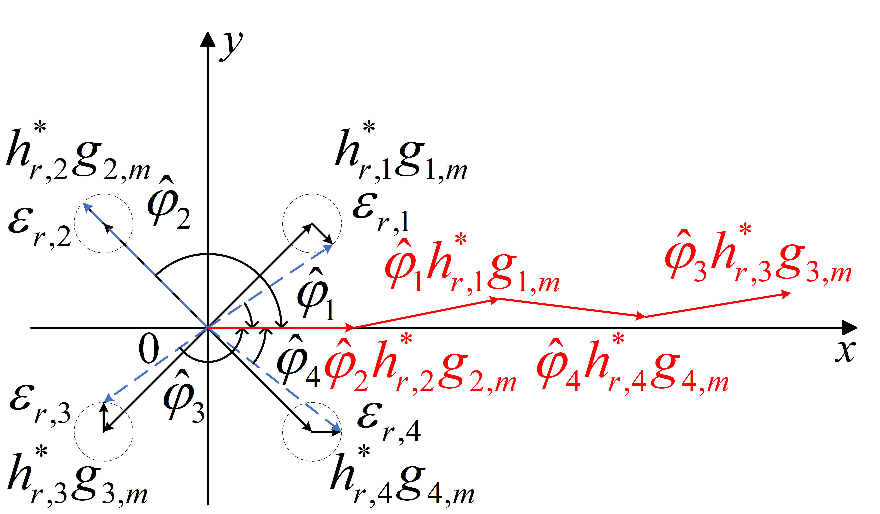}
}
\caption{ Conventional channel estimation and passive beamforming scheme. (Red arrow: cascaded channel including the RC configuration) } 
\label{fig_used} 
\end{figure}

Next, we provide some useful insights into $\emph{Proposition 1}$ by considering four special cases:
\begin{itemize}
\item[1)] As the Rician factor $F_r \to 0$, the RIS-user channel turns into the Rayleigh channel. In this case, the proposed scheme degenerates to the random phase-shift codebook scheme \cite{r8} and the received power $P_r \to P_d \beta_r \beta_g M N \left(\text{log}Q+C\right)$.
\item[2)] As the Rician factor $F_r$ approaches infinity, the RIS-user channel turns into the deterministic LoS channel, which is perfectly aligned according to the statistical CSI. In this case, we have $P_r \to P_d \beta_r \beta_g M N^2$ which characterizes the quadratic power scaling law versus the number of RIS elements\cite{r19}.
\item[3)] When considering the training overhead of $Q=1$, the environment-aware algorithm reduces to the statistical CSI-based scheme \cite{r13}. In particular, if $F_r \to 0$, we have $P_r = P_d \beta_r \beta_g M N C$, which is equivalent to a random phase shift configuration scenario. If $F_r \to +\infty$, we can achieve the power scaling law relying on the optimal RIS configuration, i.e. $P_r = P_d \beta_r \beta_g M N^2$.
\item[4)] When the training overhead reaches the maximum overhead $Q_{\text{max}}=2^{bN}$, we arrive at the quadratic scaling law of the average received power at the user with respect to the number of RIS elements $N$ \cite{r8}, which is comparable to the optimal RIS configuration.
\end{itemize}

As such, the proposed environment-aware codebook scheme strikes favorable tradeoffs between the statistical CSI-based RIS configuration and the optimal RIS configuration method by flexibly adapting the codebook size $Q$.

\subsection{Imperfect CSI}
\begin{figure*}[!t] 
\centering 
\subfloat[\scriptsize Perfect CSI scenario. (Red arrow: cascaded channel including the RC configuration)] 
{
\label{fig_training:subfig1}\includegraphics[width=0.95\textwidth]{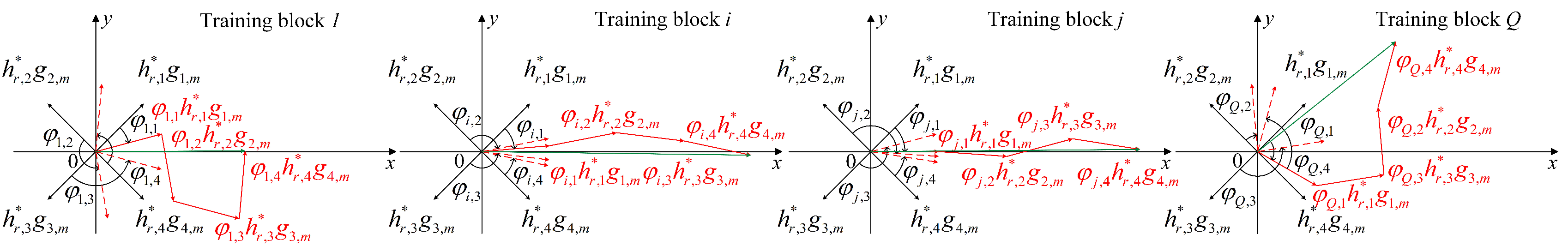} 
}

\subfloat[\scriptsize Imperfect CSI scenario. (Green arrow: composite end-to-end channel; Blue dotted arrow: composite channel with error)] 
{
\label{fig_training:subfig2}\includegraphics[width=0.95\textwidth]{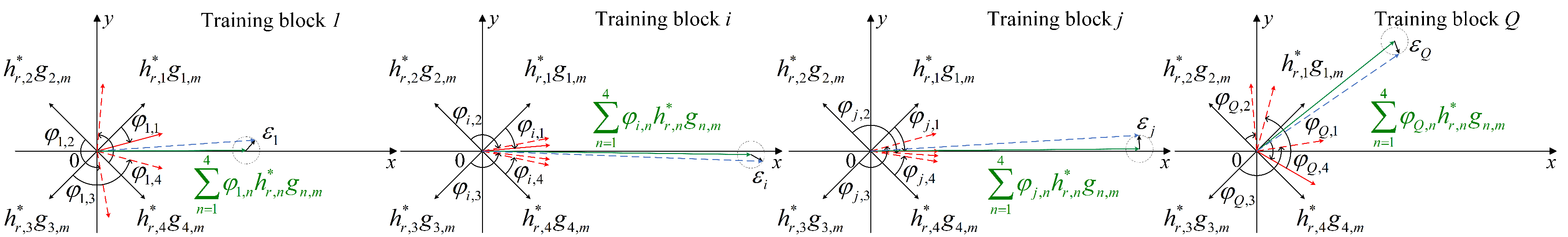}
}
\caption{ The proposed channel training-based protocol. } 
\label{fig_training} 
\end{figure*}

In this subsection, we analyze the theoretical achievable rate while considering the imperfect CSI. As shown in Fig. \ref{fig_used} and Fig. \ref{fig_training}, we provide a scenario of four RIS elements to show the effects of the channel estimation errors.

Firstly, Fig. \ref{fig_used} shows the influence of channel estimation errors on conventional channel estimation and passive beamforming schemes. Observe from Fig. \ref{fig_used}(a) that all four cascaded channels with optimal RC configuration, which are denoted by solid red arrows, are aligned when there are no channel estimation errors. In this context, the optimal achievable rate is shown in Eq. (30) of \cite{r19}. When taking into account the practical channel estimation errors, we can observe from Fig. \ref{fig_used}(b) that the optimal phase shift associated with each RIS element is rotated, which results in the deviation from the perfect case. This results in a loss of achievable rate characterized by Eq. (29) of \cite{r9}.

By contrast, the influence of channel estimation errors on the proposed environment-aware codebook scheme is shown in Fig. \ref{fig_training}, where Fig. \ref{fig_training}(a) represents four training blocks in the perfect CSI scenario and the corresponding cases in the imperfect CSI scenario are shown in Fig. \ref{fig_training}(b). 

Specifically, as shown in Fig. \ref{fig_training}, $\varphi_{q,n}, q = 1, \cdots, Q, n = 1, 2, 3, 4$ represents the RC of the $n$th RIS element at the $q$th training block. They rotate the cascaded channels with the corresponding angle to get the results shown by the red arrows. In this case, the achievable rate maximization problem is equivalent to the problem \begin{align} \label{Y:1} \max_{q \in \mathcal{Q}} Y_q = \left| \sum_{n=1}^{N} h_{r,n}^* \varphi_{q,n} g_{n,m} \right|^2. \end{align} We calculate $Y_q$ for all $Q$ training blocks and obtain the optimal index $i$ whose theoretical performance is analysed in $\emph{Proposition 1}$.

When considering the practical channel estimation errors, it is important to note that for our channel training protocol, the effect of channel estimation errors is not reflected separately at each phase shift. Instead, it distorts the composite end-to-end channels, which is shown by the green arrow. Based on the estimated composite channels (i.e., the dotted blue arrow in Fig. \ref{fig_training}(b)), maximizing the achievable rate problem is equivalent to \begin{align} \label{Y:2} \max_{q \in \mathcal{Q}} \hat{Y}_q = \left| \sum_{n=1}^{N} h_{r,n}^* \varphi_{q,n} g_{n,m} + \varepsilon_q \right|^2, \end{align} where $\varepsilon_q \sim \mathcal{CN} \left(0, \sigma_q^2 \right)$ denotes the estimation error of the composite channel in the $q$th training block. It is noted that the optimal codeword index may not be correctly selected when considering channel estimation errors. In the proposed codebook scheme, the influence of the channel estimation errors is implicitly reflected in the selection of the codeword index, whose impacts are difficult to be accurately characterized. In this paper, a tight upper bound of the achievable rate is derived in $\emph{Proposition 2}$.

$\emph{Proposition 2}$: Assume $h_{r,n}$ following the Rician channel model with Rician factor of $F_r$ , $n = 1, 2, \cdots, N$. $\varepsilon_q \sim \mathcal{CN} \left(0,\sigma_q^2 \right)$, $q = 1,2,\cdots,Q$ is the composite channel estimation error during the $q$th training block. For $N \gg 1$, the average signal power received by the user is given by \begin{align} \label{eq:25} P_r & = P_d M \beta_r \beta_g \left( N^2 F_1^2 + N F_1 F_2 \sqrt{\pi} \right. \notag \\ & \left. + N F_2^2 \frac{N + \frac{\pi}{2} (N-1) \sqrt{\frac{\beta_r \beta_g}{(N-1)\beta_r\beta_g + \sigma_q^2}} }{N+\frac{\pi}{2} \sqrt{N-1}} \left( \text{log}Q + C \right) \right) .\end{align} 

$\emph{Proof}$: Please refer to Appendix B.
\begin{figure}[!t]
\centering
\includegraphics[width=10cm]{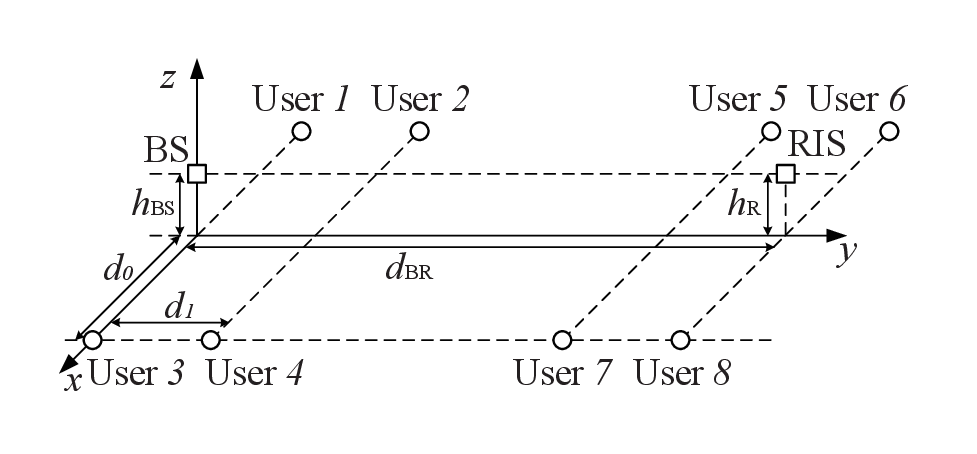}
\caption{ Schematic of an RIS-aided multi-user communication scenario.}
\label{fig_4}
\end{figure} 
\section{Simulation Results}
In this section, we provide numerical simulation results to validate the performance of the proposed scheme. All results are obtained by averaging 1,000 independent experiments. As shown in Fig. \ref{fig_4}, we consider an RIS-aided multi-user system. Specifically, the antenna array at the BS is modeled by a ULA and deployed on the $z$-axis with antenna spacing of $d_{\text{BS}} = \lambda/2$. The number of the BS antennas is set to $M = 8$. The RIS is modeled by a UPA and deployed on the $y\text{-}z$ plane with $10 \times 10$ array structure ($N = 100$) with element spacing of $d_{\text{R}} = \lambda/8$. Moreover, the Rician factor of the BS-RIS, RIS-user, and BS-user channels are $F_g = \SI{4}{dB}$, $F_r = \SI{3}{dB}$, and $F_d = \SI{-3}{dB}$, respectively. The quantization number of the discrete phase shift set is set to $b = 1$. The distance from BS to RIS, from each user to the BS-RIS line and between two adjacent users are set to $d_{\text{BR}} = \SI{100}{m}$, $d_0 = \SI{2}{m}$ and $d_1 = \SI{10}{m}$, respectively. Both the BS and RIS are situated at a height of $h_{\text{BS}} = h_{\text{R}} = \SI{5}{m}$.

The path loss of each channel is modeled as \cite{r15} \begin{align} \beta = C_0(d/d_m)^{-\alpha}, \end{align} where $d$ is the distance of the corresponding link, $C_0 = \SI{-20}{dB}$ denotes the path loss of the reference distance $d_m = \SI{1}{m}$ and $\alpha$ denotes the path loss factor. 

We assume that the path loss factor of BS-RIS, RIS-user and BS-user links are $\alpha_g = 2.4$, $\alpha_r = 2.5$ and $\alpha_d = 3.5$, respectively. Moreover, we consider the transmit power at the BS is set to $P_d = \SI{40}{dBm}$, the average power of the pilot signals is $P_{\text{ul}} = \SI{-23}{dBm}$, the average noise power at the BS and the user are $\sigma_z^2 = \SI{-110}{dBm}$ and $\sigma_k^2 = \SI{-90}{dBm}$, respectively.

\begin{figure}[!t]
\centering
\includegraphics[width=9cm]{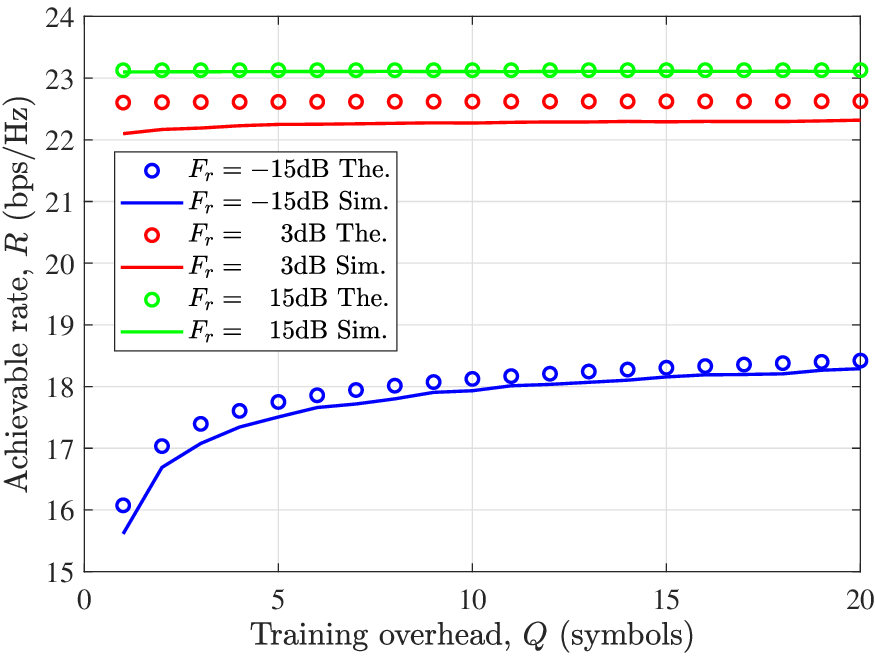}
\caption{ Achievable rate $R$ of the single user versus training overhead $Q$, where we consider perfect CSI (The.: theoretical result; Sim.: simulation result).}
\label{fig_3}
\end{figure} 

Firstly, we verify the performance analysis in a single user scenario by only considering User $8$ in Fig. \ref{fig_4}. As shown in Fig. \ref{fig_3}, we verify our analytical results in Section IV-A under different RIS-user channel conditions, where we set $F_r = \SI{-15}{dB}$, $\SI{3}{dB}$, and $\SI{15}{dB}$, respectively. It can be seen that the proposed codebook scheme gains an improved achievable rate as the codebook size increases. Moreover, the theoretical result offers a tight upper bound of the simulation result when confronting with diverse channel conditions.

\begin{figure}[!t]
\centering
\includegraphics[width=9cm]{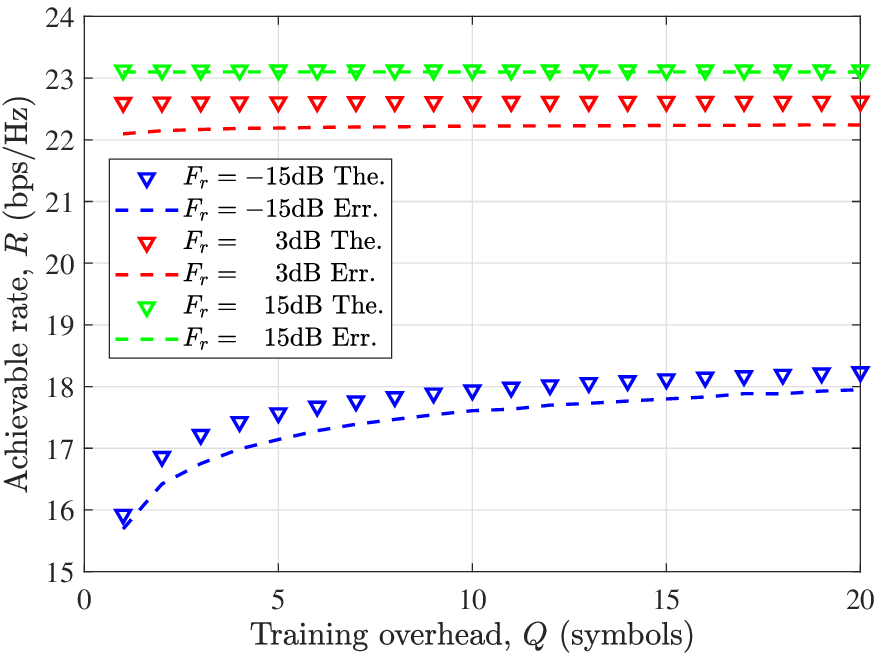}
\caption{ Achievable rate $R$ of the single user versus training overhead $Q$, where we consider imperfect CSI (Err.: simulation result with imperfect CSI).}
\label{fig_3_e}
\end{figure} 

Moreover, Fig. \ref{fig_3_e} verifies our analytical results in Section IV-B considering the channel estimation errors. It can be seen that the theoretical bound accurately characterizes the environment-aware codebook scheme for a high Rician factor, e.g., $F_r = \SI{15}{dB}$. Besides, in the case of $F_r = \SI{3}{dB}$ and $\SI{-15}{dB}$, the theoretical results provide a tight upper bound. Compared to Fig. \ref{fig_3}, the performance degradation at $F_r = \SI{15}{dB}$ is negligible, which implies that the codebook based on statistical CSI achieves optimal performance at high Rician factors. At $F_r = \SI{3}{dB}$ and $\SI{-15}{dB}$, there exists moderate performance degradation in the proposed scheme, indicating strong robustness against noise.

\begin{figure}[!t]
\centering
\includegraphics[width=9cm]{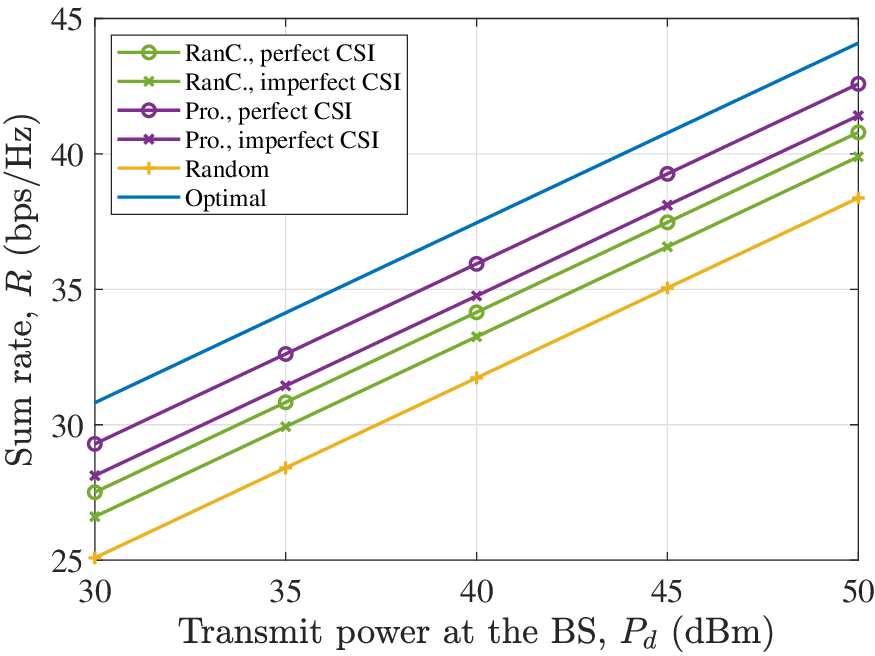}
\caption{ Sum rate $R$ versus transmit power $P_d$ at the BS (RanC.: random codebook; Pro.: proposed scheme; Random: random configuration; Optimal: optimal configuration).}
\label{fig_5}
\end{figure}

Next, as shown in Fig. \ref{fig_5}, we compare the sum rate versus total transmit power at BS while activating Users $6$ and $8$. The Rician factor of the RIS-user link is set to $F_r = \SI{10}{dB}$. The training overhead is set to $Q = 100$. Note that the sum rate is improved as the transmit power increases, and exhibits a logarithmic growth trend. In addition, the proposed scheme outperforms the random codebook scheme in both the perfect and imperfect CSI scenarios. Specifically, a sum rate gain of $\SI{1.78}{bps/Hz}$ is observed under perfect CSI. This is because the proposed scheme generates the RIS RC codebook by leveraging the statistical CSI.
\begin{figure}[!t] 
\centering 
\subfloat[\scriptsize $\text{User}\ k, k=5,6,7,8$.] 
{
\label{fig_6:subfig1}\includegraphics[width=0.3\textwidth]{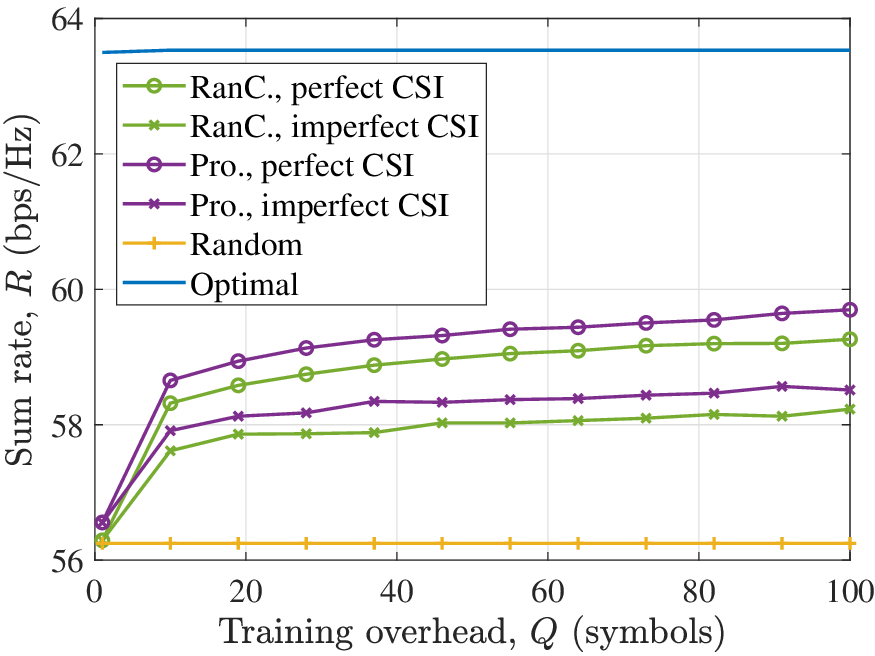}
}
\subfloat[\scriptsize $\text{User}\ k, k=2,4,6,8$.]
{
\label{fig_6:subfig2}\includegraphics[width=0.3\textwidth]{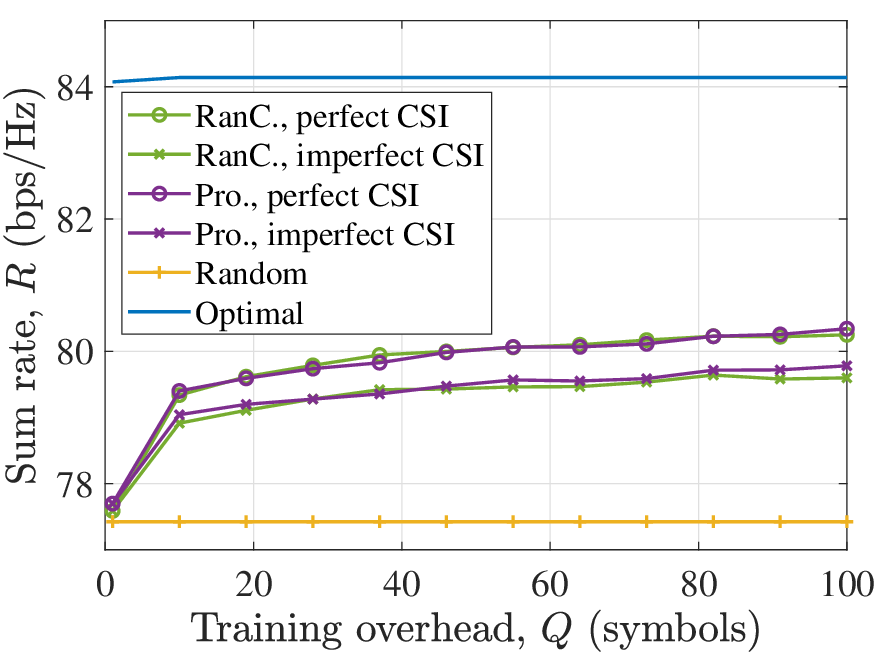}
}

\subfloat[\scriptsize $\text{User}\ k, k=2,4,5,6,7,8$.] 
{
\label{fig_6:subfig3}\includegraphics[width=0.3\textwidth]{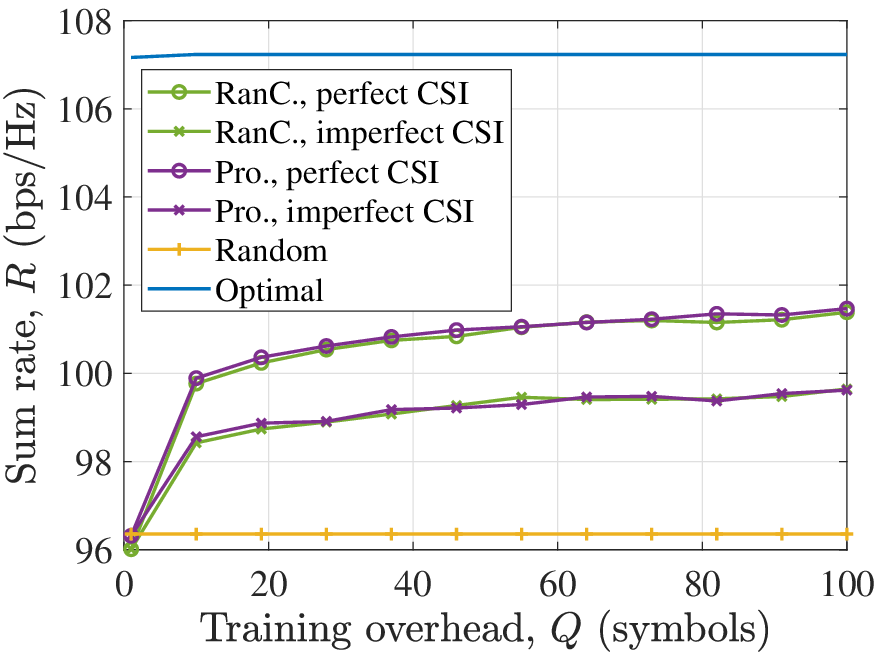}
}
\subfloat[\scriptsize $\text{User}\ k, k=6,8$.]
{
\label{fig_6:subfig4}\includegraphics[width=0.3\textwidth]{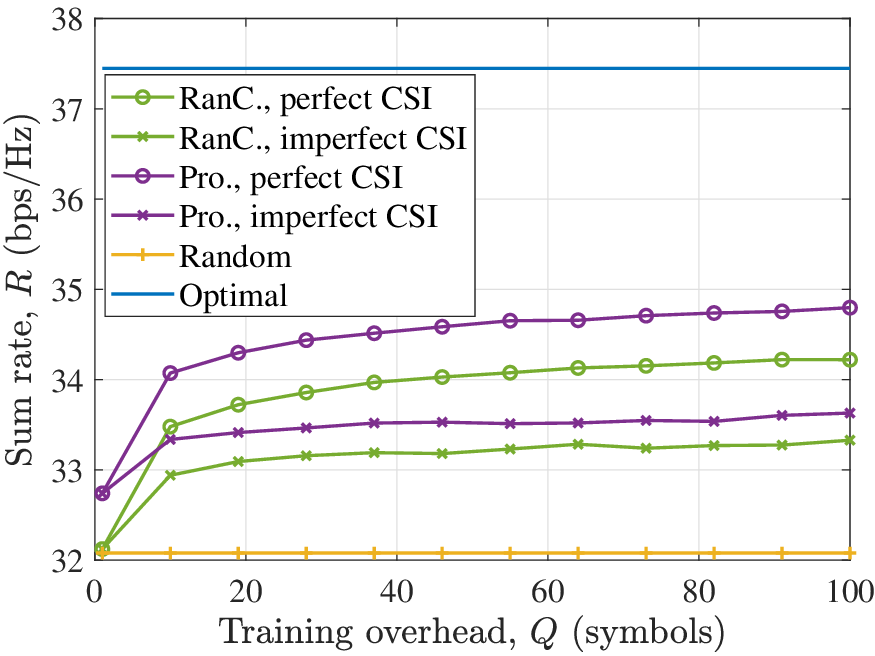}
}
\caption{ Sum rate $R$ versus training overhead $Q$.} 
\label{fig_6} 
\end{figure}

Next, we evaluate the robustness of the proposed environment-aware codebook-based scheme by considering four different setups in Fig. \ref{fig_6}. Note that under all setups, the achievable sum rate increases with the training overhead. Besides, observe from Fig. \ref{fig_6} that the proposed scheme always outperforms the random codebook scheme corresponding to the training overhead $Q = 1$, due to the fact that the proposed scheme utilizes the statistical CSI to design the phase shift codebook. Further comparing Figs. \ref{fig_6}(a)-\ref{fig_6}(c) that activating more users results in an improved sum rate, thanks to the multi-user multiplexing gain. For instance, the sum rate increases by \SI{24.90}{bps/Hz} from activating users $6$ and $8$ to activating users $5, 6, 7$ and $8$. Furthermore, we note that the advantage of the proposed scheme over the random codebook scheme is more significant in Figs. \ref{fig_6}(a) and \ref{fig_6}(d). Both these two cases correspond to activating users near the RIS. Moreover, Fig. \ref{fig_6} shows that the influence caused by the channel estimation errors becomes severe as the training overhead increases, which is consistent with our analysis accounting for the imperfect CSI in Section IV-B.

\begin{figure}[!t]
\centering
\includegraphics[width=9cm]{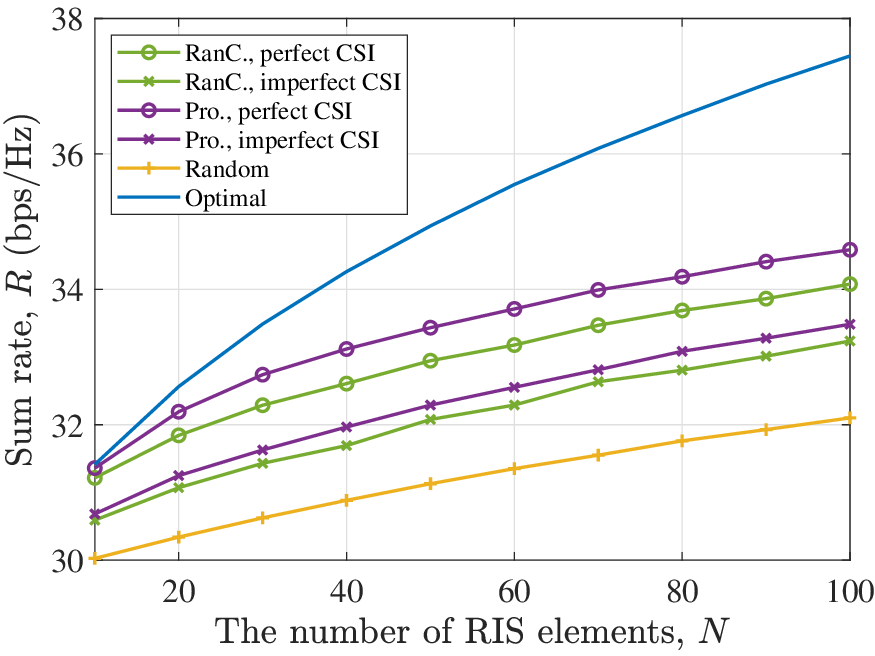}
\caption{ Sum rate $R$ versus the number of RIS elements $N$.}
\label{fig_7}
\end{figure} 

Next, Fig. \ref{fig_7} compares the sum rate versus the number of RIS elements, considering users $6$ and $8$ and setting the training overhead to $Q = 50$. It can be observed that the proposed scheme outperforms the random codebook scheme under all setups. Besides, as the number of RIS elements increases from $N=10$ to $N=100$, the proposed scheme results in a more significant sum rate improvement with diminishing return, which is due to the fact that a larger number of reflecting elements could reap more radiating energy by coherently superimposing all cascaded links.

\begin{figure}[!t]
\centering
\includegraphics[width=9cm]{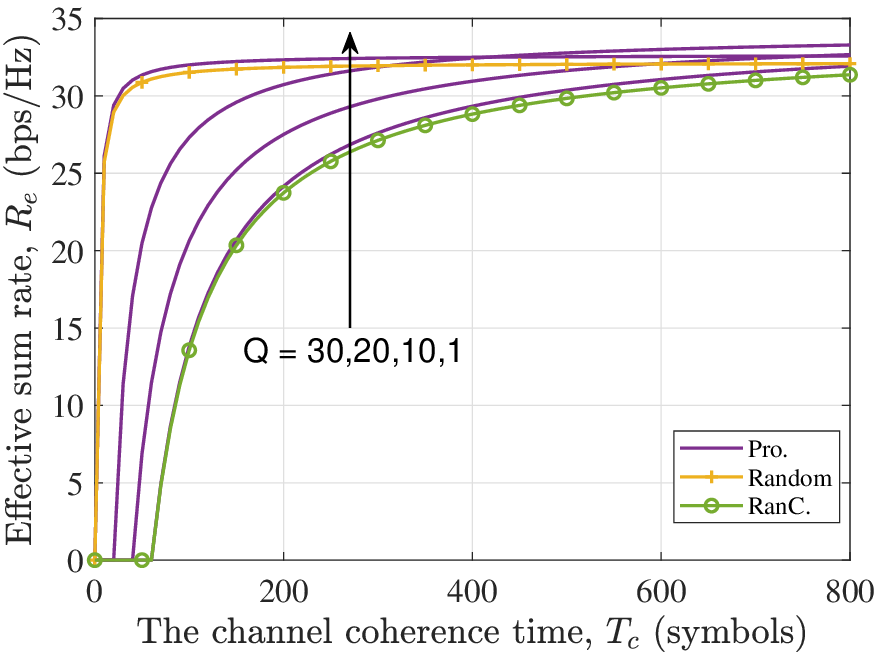}
\caption{The effective sum rate $R_e$ versus the channel coherence time $T_c$.}
\label{fig_R_T}
\end{figure} 

Finally, Fig. \ref{fig_R_T} evaluates the effective sum rate versus the channel coherence time, where we only activate users $6$ and $8$ and consider four different training overhead scenarios. Here, the effective sum rate is defined by \begin{align} R_e = \frac{T_c-\tau}{T_c} \mathbb{E} \left\{ \sum_{k=1}^{K} \text{log}_2 \left( 1 + \frac{\left| \mathbf{h}_{k}^H \mathbf{w}_{k} \right|^2}{\sum_{l \neq k}^K \left| \mathbf{h}_{k}^H \mathbf{w}_{l} \right|^2 + \sigma_k^2} \right) \right\}, \end{align} where $T_c$ is the channel coherence time and $\tau$ is the pilot overhead. It can be seen that the proposed scheme can flexibly adapt the pilot overhead in the face of different values of channel coherence time. Specifically, for a rapidly time-varying channel (e.g., $T_c \leq 200$), the effective sum rate of the proposed scheme with training overhead $Q = 1$ attains the best performance. However, as the channel coherence time becomes longer, the increased training overhead can lead to a higher effective sum rate. Therefore, based on the specific coherence time, the proposed scheme can flexibly adjust the training overhead, thus maximizing the effective sum rate.

\section{Conclusion}
In this paper, we have proposed a channel training-based protocol for RIS-assisted MU-MISO systems, where an environment-aware RC codebook is designed. Specifically, during the off-line stage, the proposed scheme generates a set of virtual channels based on the statistical CSI and obtains an environment-aware codebook according to the off-line AO algorithm. Then, during the on-line stage, we have configured the RIS based on the environment-aware codebook and have directly estimated the composite end-to-end channels. Subsequently, we have performed the ZF precoding design at the BS to obtain candidate channels. Finally, the optimal codeword resulting in the highest sum rate has been adopted to configure the RIS for assisting in the data transmission. Furthermore, we have analyzed the performance of the proposed scheme accounting for both the perfect and the imperfect CSI scenarios and have evaluated the achievable rate of the environment-aware codebook scheme through simulation results. It has been shown that the proposed scheme has better performance than traditional codebook schemes, and can strike a favorable trade-off between performance and the training overhead. We highlight that the proposed scheme can be applied to a wider range of wireless systems including cell-free massive MIMO systems and OFDM systems by leveraging the statistical CSI therein \cite{TVT_2024_An_Adjustable, IoTJ_2024_Xu_Algorithm, shi2022spatially,add21}.

\appendices
\section{Proof of Proposition 1}
By applying the assumptions of $\emph{Proposition 1}$, let $\mathcal{Q} = \left\{1, 2, \cdots, Q \right\}$, the average received power $P_r$ at the single user with codebook based RC configuration can be expressed as 
\begin{align} \label{eq:P_r} P_r = P_d \mathbb{E} \left \{ \max_{q \in \mathcal{Q}} \lVert \mathbf{h}_r^H \mathbf{\Phi}_q \mathbf{G} \rVert^2 \right\}. \end{align}

Due to the fact that the maximum value of the sum does not exceed the sum of the maximum values, (\ref{eq:P_r}) can be further written as 
\begin{align} P_r & = P_d \mathbb{E} \left\{ \max_{q \in \mathcal{Q}} \left\lVert \sum_{n=1}^{N} h_{r,n}^* \varphi_{q,n} \mathbf{g}_n^T \right\rVert^2 \right\} \notag\\ & \le P_d \mathbb{E} \left \{ \sum_{m=1}^{M} \max_{q \in \mathcal{Q}} \left| \sum_{n=1}^{N} h_{r,n}^* \varphi_{q,n} g_{n,m} \right|^2 \right\} \notag\\ &= P_d \beta_r \beta_g M \mathbb{E} \left\{\max_{q \in \mathcal{Q}} \left|\sum_{n=1}^{N} \overline{h}_{r,n}^* \varphi_{q,n} \overline{g}_{n,m} \right|^2 \right\}, \end{align} 
where $\mathbf{g}_n^T$ is the $n$th row of $\mathbf{G}$ and ‘$\le$’ reduces to ‘=’ if and only if $\text{rank}(\mathbf{G}) = 1$; $h_{r,n}$ and $g_{n,m}$ are the $n$th entry of $\mathbf{h}_r$ and $\mathbf{g}_m$, respectively; $\overline{h}_{r,n}^*$ and $\overline{g}_{n,m}$ denote their normalized results, respectively. $\varphi_{q,n}$ is the $n$th entry on the diagonal of $\mathbf{\Phi}_q$. 

Since we have $\overline{h}_{r,n}^* = F_1 \left(h_{r,n}^{\text{LoS}}\right)^* + F_2 \left(h_{r,n}^{\text{NLoS}}\right)^*$ and $\overline{g}_{n,m} = g_{n,m}^{\text{LoS}}$, $\left|\sum_{n=1}^{N} \overline{h}_{r,n}^* \varphi_{q,n} \overline{g}_{n,m}\right|^2$ can be further written as
\begin{align} \label{eq:29} & \left|\sum_{n=1}^{N} \overline{h}_{r,n}^* \varphi_{q,n} \overline{g}_{n,m}\right|^2 \notag\\ = &\left|\sum_{n=1}^{N} \left(F_1 \left(h_{r,n}^{\text{LoS}}\right)^* + F_2 \left(h_{r,n}^{\text{NLoS}}\right)^* \right) \varphi_{q,n} g_{n,m}^{\text{LoS}}\right|^2 \notag\\ = &\left|\sum_{n=1}^{N} F_1 \left(h_{r,n}^{\text{LoS}}\right)^* \varphi_{q,n} g_{n,m}^{\text{LoS}}\right|^2+ \left|\sum_{n=1}^{N} F_2 \left(h_{r,n}^{\text{NLoS}}\right)^* \varphi_{q,n} g_{n,m}^{\text{LoS}}\right|^2 \notag\\ & + 2 \Re \left \{ \left(\sum_{n=1}^{N} F_1 \left(h_{r,n}^{\text{LoS}}\right)^* \varphi_{q,n} g_{n,m}^{\text{LoS}}\right)^* \right. \notag \\ & \left. \times \left(\sum_{n=1}^{N} F_2 \left(h_{r,n}^{\text{NLoS}}\right)^* \varphi_{q,n} g_{n,m}^{\text{LoS}}\right) \right \}. \end{align} 

Note that we adopt the environment-aware scheme relying on the statistical CSI. In each channel training block, we generate the NLoS component of the RIS-user channel and configure the phase shift of the $n$th RIS element as \begin{align} \label{phase_shift_without_error} \theta_{q,n} = \angle \left( \sqrt{F_r} h_{r,n}^{\text{LoS}} + \tilde{h}_{r,q,n}^{\text{NLoS}} \right) - \angle \left( g_{n,m}^{\text{LoS}} \right),\end{align} where $\tilde{h}_{r,q,n}^{\text{NLoS}}$ is the random NLoS component generated off-line. Thus when considering the Rayleigh fading in the proposed scheme, it is equivalent to adopting a random phase-shift configuration codebook.

Then, we consider the three terms of (\ref{eq:29}) separately. Specifically, for the first entry of (\ref{eq:29}), the LoS component of multiple reflected channels is largely aligned thus it can be expressed as
\begin{align} \label{eq:LoS_withouterror} & \mathbb{E} \left \{ \max_{q \in \mathcal{Q}} \left|\sum_{n=1}^{N} F_1 \left(h_{r,n}^{\text{LoS}}\right)^* \varphi_{q,n} g_{n,m}^{\text{LoS}}\right|^2 \right\} \notag\\ &= F_1^2 \left | \sum_{n=1}^{N} \left| \left(h_{r,n}^{\text{LoS}}\right)^* \right| \left|g_{n,m}^{\text{LoS}}\right| \right |^2 = F_1^2 N^2. \end{align} 

The other two entries of (\ref{eq:29}) are related to the NLoS component, where we can obtain $\sum_{n=1}^{N} \varphi_{q,n} \left(h_{r,n}^{\text{NLoS}}\right)^* g_{n,m}^{\text{LoS}} \sim \mathcal{CN} \left(0, N\right)$ as $N \to \infty$ based on Lindeberg-Levy central limit theorem \cite{r19}. 

Specifically, the second entry reflects the effects on the NLoS component of the channel. According to \emph{Lemma 1} in \cite{r8}, we have \begin{align}\max_{q \in \mathcal{Q}} \left| \sum_{n=1}^{N} \varphi_{q,n} \left(h_{r,n}^{\text{NLoS}}\right)^* g_{n,m}^{\text{LoS}} \right|^2 \overset{d}{=} N \sum_{j=1}^{Q} \frac{1}{Q-j+1} \rho_j, \end{align} where $\rho_j$ are i.i.d. standard exponential random variables with a rate parameter of $1$. Thus we have $\mathbb{E} \left\{ \rho_j\right\} = 1$ and it can be obtained that 
\begin{align} \label{eq:NLoS_withouterror} & \mathbb{E} \left\{ \max_{q \in \mathcal{Q}} \left|\sum_{n=1}^{N} F_2 \left(h_{r,n}^{\text{NLoS}}\right)^* \varphi_{q,n} g_{n,m}^{\text{LoS}}\right|^2 \right\} \notag\\ =& F_2^2 \mathbb{E} \left\{N \sum_{j=1}^{Q} \frac{1}{Q-j+1} \rho_j \right\} \notag\\ =& F_2^2 N\left(\text{log}Q+C\right). \end{align} 

Since we have $\mathbb{E} \left\{ \sum_{n=1}^{N} \varphi_{q,n} \left(h_{r,n}^{\text{NLoS}}\right)^* g_{n,m}^{\text{LoS}} \right\} = 0$, the third entry of (\ref{eq:29}) can be further derived as 
\begin{align} \label{eq:33} & \mathbb{E} \left\{ \max_{q \in \mathcal{Q}} 2 \Re \left\{ \left(\sum_{n=1}^{N} F_1 \left(h_{r,n}^{\text{LoS}}\right)^* \varphi_{q,n} g_{n,m}^{\text{LoS}}\right)^* \right. \right. \notag\\ & \left. \left. \times \left(\sum_{n=1}^{N} F_2 \left(h_{r,n}^{\text{NLoS}}\right)^* \varphi_{q,n} g_{n,m}^{\text{LoS}}\right) \right\} \right\} \notag\\ = &2 F_1 F_2 \Re \left\{ \mathbb{E} \left\{ \max_{q \in \mathcal{Q}} \left\{ \sum_{n=1}^{N} \left|\varphi_{q,n}\right|^2 h_{r,n}^{\text{LoS}} \left(h_{r,n}^{\text{NLoS}}\right)^* \left|g_{n,m}^{\text{LoS}}\right|^2 + \right. \right. \right. \notag\\ & \left. \left. \left. \left( \sum_{i=1}^{N} \varphi_{q,i} \left(h_{r,i}^{\text{NLoS}}\right)^* g_{i,m}^{\text{LoS}} \right) \left( \sum_{j \neq i}^{N} \varphi_{q,j}^* h_{r,j}^{\text{LoS}} (g_{j,m}^{\text{LoS}})^* \right) \right\} \right\} \right\} \notag\\ \overset{(a)}{=} &2 F_1 F_2 \Re \left \{ \mathbb{E} \left\{ \sum_{n=1}^{N} h_{r,n}^{\text{LoS}} \left(h_{r,n}^{\text{NLoS}}\right)^* \right\} \right\} \notag\\ \overset{(b)}{<} &2 F_1 F_2 \sum_{n=1}^{N} \left( \left|h_{r,n}^{\text{LoS}}\right| \mathbb{E} \left\{ \left|(h_{r,n}^{\text{NLoS}})^*\right| \right\} \right) \notag\\ \overset{(c)}{=} &\sqrt{\pi} F_1 F_2 N, \end{align} 
where $(a)$ holds because we have $\mathbb{E} \left\{\sum_{i=1}^{N} \varphi_{q,i} \left(h_{r,i}^{\text{NLoS}}\right)^* g_{i,m}^{\text{LoS}}\right\}=0$, $\left|g_{n,m}^{\text{LoS}}\right|=1$ and $\left|\varphi_{q,n}\right|=1$; $(b)$ holds as we have $\Re \left\{\mathbb{E} \left\{ \sum_{n=1}^{N} h_{r,n}^{\text{LoS}} \left(h_{r,n}^{\text{NLoS}}\right)^* \right\} \right\} < \sum_{n=1}^{N} \mathbb{E}\left\{ \Re \left\{ h_{r,n}^{\text{LoS}} \right\} \right\} \mathbb{E}\left\{ \Re \left\{ \left(h_{r,n}^{\text{NLoS}}\right)^* \right\} \right\}$, $\Re \left\{ h_{r,n}^{\text{LoS}} \right\} \leq \left|h_{r,n}^{\text{LoS}}\right|$ and $\Re \left\{ \left(h_{r,n}^{\text{NLoS}}\right)^* \right\} \leq \left|(h_{r,n}^{\text{NLoS}})^*\right|$; Moreover, equality $(c)$ holds due to the fact that $\left|\left(h_{r,n}^{\text{NLoS}}\right)^*\right|$ follows Rayleigh distribution with mean value of $\sqrt{\pi}/2$.

Substituting (\ref{eq:LoS_withouterror}), (\ref{eq:NLoS_withouterror}) and (\ref{eq:33}) into (\ref{eq:29}), the proof is completed.$\hfill\blacksquare$

\section{Proof of Proposition 2}
Similar to (\ref{eq:29}), the average received power $P_r$ of the single user when considering channel estimation errors is given by \begin{align} \label{eq:34} P_r & = P_d \beta_r \beta_g M \mathbb{E} \left\{ \left|\sum_{n=1}^{N} F_1 \left(h_{r,n}^{\text{LoS}}\right)^* \varphi_{\hat{q},n} g_{n,m}^{\text{LoS}}\right|^2 \right. \notag\\ & \left. + \left|\sum_{n=1}^{N} F_2 \left(h_{r,n}^{\text{NLoS}}\right)^* \varphi_{\hat{q},n} g_{n,m}^{\text{LoS}}\right|^2 \right. \notag\\ & \left. + 2 \Re \left \{ \left(\sum_{n=1}^{N} F_1 \left(h_{r,n}^{\text{LoS}}\right)^* \varphi_{\hat{q},n} g_{n,m}^{\text{LoS}}\right)^* \right. \right. \notag\\ & \left. \left. \times \left(\sum_{n=1}^{N} F_2 \left(h_{r,n}^{\text{NLoS}}\right)^* \varphi_{\hat{q},n} g_{n,m}^{\text{LoS}}\right) \right \} \right\}, \end{align} where $\varphi_{\hat{q},n}$ is the RC with selected index $\hat{q}$ for the $n$th RIS element in the presence of channel estimation errors, which is obtained by \begin{align} \label{eq:36} \varphi_{\hat{q},n} = \arg \max_{q \in \mathcal{Q}} \left| \sum_{i\neq n}^{N} h_{r,i}^* \varphi_{q,i} g_{i,m} + h_{r,n}^* \varphi_{q,n} g_{n,m} + \varepsilon_q \right|^2 ,\end{align} where $\varphi_{q,n}$ represents the RC of the $n$th RIS element at the $q$th training block, and $\varphi_{q,i}, i \neq n$ represent the fixed ($N-1$) RCs when optimizing $\varphi_{q,n}$. $\varepsilon_q \sim \mathcal{CN} \left( 0, \sigma_q^2 \right)$ is the composite channel estimation error in the $q$th training block.

Then, as illustrated in Fig. \ref{fig_training}, we select the optimal RC configuration maximizing the sum rate in the imperfect CSI scenario. Since the codewords are obtained based on the statistical CSI, the first entry of (\ref{eq:34}) is equal to that without channel estimation errors. Moreover, as we have $\mathbb{E} \left\{\sum_{i=1}^{N} \left(h_{r,n}^{\text{NLoS}} \right)^* \varphi_{\hat{q},i} g_{i,m}^{\text{LoS}}\right\} = 0$, the result of the third part remains the same with that considering perfect CSI, i.e., (\ref{eq:33}).

Under imperfect CSI conditions, the second entry of (\ref{eq:34}) still follows an exponential distribution at each training block. However, the $Q$th order statistics is a scaling version of that in (32). Let $u$ denote the scaling factor. Specifically, we first consider the effect of the channel estimation errors on a single RC, keeping the remaining ($N-1$) terms fixed, i.e., $\sum_{i \neq n}^{N} \sqrt{\beta_r} (h_{r,i}^{\text{NLoS}})^* \varphi_{q,i} \sqrt{\beta_g} g_{i,m}^{\text{LoS}} \sim \mathcal{CN} \left(0, (N-1)\beta_r\beta_g\right)$. Next, we derive the scaling factor to characterize the influence of the channel estimation errors. 

Let $h_{c,n}^* = \sqrt{\beta_r}(h_{r,n}^{\text{NLoS}})^* \sqrt{\beta_g}g_{n,m}^{\text{LoS}}$ represent the NLoS component of the cascaded channel spanning from the user to the BS through the $n$th RIS element. Hence, by solving (\ref{Y:1}), it is straightforward to obtain the optimal RC for the $n$th RIS element $\bar{\varphi}_{q,n} = \frac{h_{c,n}}{\left| h_{c,n}^* \right|} \frac{\sum_{i \neq n}^{N} h_{c,i}^* \varphi_{q,i} }{\left| \sum_{i \neq n}^{N} h_{c,i}^* \varphi_{q,i} \right|}$, and the second term in (\ref{eq:34}) can be simplified to \begin{align} u_{o} &\triangleq \left|\sum_{i \neq n}^{N} h_{c,i}^* \varphi_{q,i} + h_{c,n}^* \bar{\varphi}_{q,n} \right|^2 \notag \\ & = \left| \sum_{i \neq n}^{N} h_{c,i}^* \varphi_{q,i} \right|^2 + \left| h_{c,n} \right|^2 + 2 \Re\left\{ \left| \sum_{i \neq n}^{N} h_{c,i} \varphi_{q,i}^*\right| \left| h_{c,n}^* \right| \right\}.\end{align}

Moreover, we obtain the optimal RC under imperfect CSI by solving (\ref{Y:2}), i.e., $\hat{\varphi}_{q,n} = \frac{h_{c,n}}{\left|h_{c,n}^* \right|} \frac{\sum_{i \neq n}^{N} h_{c,i}^* \varphi_{q,i} + \varepsilon_q}{\left| \sum_{i \neq n}^{N} h_{c,i}^* \varphi_{q,i} + \varepsilon_q \right|}$, we have \begin{align} \label{sumrate_e} u_{e} &\triangleq \left| \sum_{i \neq n}^{N} h_{c,i}^* \varphi_{q,i} + h_{c,n}^* \hat{\varphi}_{q,n} \right|^2 \notag \\ & = \left( \left| \sum_{i \neq n}^{N} h_{c,i}^* \varphi_{q,i} \right|^2 + \left| h_{c,n}^* \right|^2 \right. \notag \\ & \left. + 2 \Re \left\{ \sum_{i \neq n}^{N} h_{c,i} \varphi_{q,i}^* \left|h_{c,n}^* \right| \frac{\sum_{i \neq n}^{N} h_{c,i}^* \varphi_{q,i} + \varepsilon_q}{\left| \sum_{i \neq n}^{N} h_{c,i}^* \varphi_{q,i} + \varepsilon_q \right|} \right\} \right).\end{align}

The expectation of the third entry in (\ref{sumrate_e}) can be expressed as \begin{align} & \mathbb{E} \left\{ 2 \Re \left\{ \sum_{i \neq n}^{N} h_{c,i} \varphi_{q,i}^* \left|h_{c,n}^* \right| \frac{\sum_{i \neq n}^{N} h_{c,i}^* \varphi_{q,i} + \varepsilon_q}{\left| \sum_{i \neq n}^{N} h_{c,i}^* \varphi_{q,i} + \varepsilon_q \right|} \right\} \right\} \notag \\ & = 2 \mathbb{E} \left\{ \left|h_{c,n}^* \right| \Re \left\{ \sum_{i \neq n}^{N} h_{c,i} \varphi_{q,i}^* \frac{\sum_{i \neq n}^{N} h_{c,i}^* \varphi_{q,i} + \varepsilon_q}{\left| \sum_{i \neq n}^{N} h_{c,i}^* \varphi_{q,i} + \varepsilon_q \right|} \right\} \right\}.\end{align}

According to the assumptions of $h_{c,n}^* \sim \mathcal{CN} \left(0, \beta_r \beta_g \right), n = 1, 2, \cdots, N$, $\sum_{i \neq n}^{N} h_{c,i} \varphi_{q,i}^* \sim \mathcal{CN} \left(0, (N-1)\beta_r \beta_g \right)$, $\varepsilon_q \sim \mathcal{CN} \left(0, \sigma_q^2 \right), q = 1, 2, \cdots, Q$, we have \begin{align} & \mathbb{E} \left\{ \left| \sum_{i \neq n}^{N} h_{c,i}^* \varphi_{q,i} \right|^2 \right\} = (N-1) \beta_r \beta_g, \notag \\ & \mathbb{E} \left\{ \left| h_{c,n}^* \right|^2 \right\} = \beta_r \beta_g, \ \mathbb{E} \left\{ \left| h_{c,n}^* \right| \right\} = \frac{\sqrt{\pi}}{2} \sqrt{\beta_r \beta_g}, \notag \\ & \mathbb{E} \left\{\Re\left\{ \left|\sum_{i \neq n}^{N} h_{c,i} \varphi_{q,i}^*\right| \left| h_{c,n}^* \right| \right\}\right\} = \frac{\pi}{4} \sqrt{N-1} \beta_r \beta_g, \notag \\ & \mathbb{E} \left\{ \Re \left\{ \sum_{i \neq n}^{N} h_{c,i} \varphi_{q,i}^* \frac{\sum_{i \neq n}^{N} h_{c,i}^* \varphi_{q,i} + \varepsilon_q}{\left| \sum_{i \neq n}^{N} h_{c,i}^* \varphi_{q,i} + \varepsilon_q \right|} \right\} \right\} \notag \\ & = \frac{\sqrt{\pi} (N-1)\beta_r\beta_g}{2 \sqrt{(N-1)\beta_r\beta_g + \sigma_q^2}}.\end{align}

As a result, the scaling factor is obtained by \begin{align} \label{eq:u} u & \triangleq \frac{\mathbb{E} \left\{ u_e \right\}}{\mathbb{E} \left\{ u_o \right\}} \notag \\ & = \frac{N + \frac{\pi}{2} (N-1) \sqrt{\frac{\beta_r \beta_g}{(N-1)\beta_r\beta_g + \sigma_q^2}} }{N+\frac{\pi}{2} \sqrt{N-1}} .\end{align} 

Applying (\ref{eq:u}) to the second entry of (\ref{eq:34}), we arrive at \begin{align} & \mathbb{E} \left\{ \left|\sum_{n=1}^{N} F_2 \left(h_{r,n}^{\text{NLoS}}\right)^* \varphi_{\hat{q},n} g_{n,m}^{\text{LoS}}\right|^2 \right\} \notag \\ & < F_2^2 u \mathbb{E} \left\{ \max_{q \in \mathcal{Q}} \left|\sum_{n=1}^{N} \left(h_{r,n}^{\text{NLoS}}\right)^* \varphi_{q,n} g_{n,m}^{\text{LoS}}\right|^2 \right\} \notag \\ & = F_2^2 \frac{N + \frac{\pi}{2} (N-1) \sqrt{\frac{\beta_r \beta_g}{(N-1)\beta_r\beta_g + \sigma_q^2}} }{N+\frac{\pi}{2} \sqrt{N-1}} N (\text{log}Q + C).\end{align}

Combining the three entries in (\ref{eq:34}), the proof is completed. $\hfill\blacksquare$

\bibliographystyle{IEEEtran}
\bibliography{IEEEabrv,environment_aware_codebook}

\vfill
\end{document}